\documentstyle[psfig]{mn}

\newif\ifAMStwofonts
\AMStwofontstrue

\newcommand{\be}{\begin{equation}}    % for lazy typers
\newcommand{\ee}{\end{equation}}
\newcommand{\beq}{\begin{eqnarray}}
\newcommand{\eeq}{\end{eqnarray}}
\def\msun{M_\odot}

\def\nn{\nonumber}

\def\SNR{\mbox{SNR}}

\def\rms{\mbox{rms}}
\def\sup{\mbox{sup}}

\def\etal{{\it et al.~}}
\def\ie{{\frenchspacing\it i.e. }}
\def\eg{{\frenchspacing\it e.g. }}
\def\gtsima{$\; \buildrel > \over \sim \;$}
\def\ltsima{$\; \buildrel < \over \sim \;$}
\def\gsim{\lower.5ex\hbox{\gtsima}}
\def\lsim{\lower.5ex\hbox{\ltsima}}
\def\lanlge{<}
\def\ranlge{>}

\ifoldfss
  \ifCUPmtlplainloaded \else
    \NewTextAlphabet{textbfit} {cmbxti10} {}
    \NewTextAlphabet{textbfss} {cmssbx10} {}
    \NewMathAlphabet{mathbfit} {cmbxti10} {} % for math mode
    \NewMathAlphabet{mathbfss} {cmssbx10} {} %  "   "    "
  \fi
  \ifAMStwofonts
    \ifCUPmtlplainloaded \else
      \NewSymbolFont{upmath} {eurm10}
      \NewSymbolFont{AMSa} {msam10}
      \NewMathSymbol{\upi}     {0}{upmath}{19}
      \NewMathSymbol{\umu}     {0}{upmath}{16}
      \NewMathSymbol{\upartial}{0}{upmath}{40}
      \NewMathSymbol{\leqslant}{3}{AMSa}{36}
      \NewMathSymbol{\geqslant}{3}{AMSa}{3E}

      \let\leq=\leqslant \let\le=\leqslant
      \let\geq=\geqslant 
    \fi
  \fi
\fi % End of OFSS

\ifnfssone
  \newmathalphabet{\mathit}
  \addtoversion{normal}{\mathit}{cmr}{m}{it}
  \addtoversion{bold}{\mathit}{cmr}{bx}{it}
  \newmathalphabet{\mathbfit} % math mode version of \textbfit{..}
  \addtoversion{normal}{\mathbfit}{cmr}{bx}{it}
  \addtoversion{bold}{\mathbfit}{cmr}{bx}{it}
  \newmathalphabet{\mathbfss} % math mode version of \textbfss{..}
  \addtoversion{normal}{\mathbfss}{cmss}{bx}{n}
  \addtoversion{bold}{\mathbfss}{cmss}{bx}{n}
  \ifAMStwofonts
    \ifCUPmtlplainloaded \else
      \UseAMStwoboldmath
      \makeatletter
      \new@mathgroup\upmath@group
      \define@mathgroup\mv@normal\upmath@group{eur}{m}{n}
      \define@mathgroup\mv@bold\upmath@group{eur}{b}{n}
      \edef\UPM{\hexnumber\upmath@group}
      \new@mathgroup\amsa@group
      \define@mathgroup\mv@normal\amsa@group{msa}{m}{n}
      \define@mathgroup\mv@bold\amsa@group{msa}{m}{n}
      \edef\AMSa{\hexnumber\amsa@group}
      \makeatother
      \mathchardef\upi="0\UPM19
      \mathchardef\umu="0\UPM16
      \mathchardef\upartial="0\UPM40
      \mathchardef\leqslant="3\AMSa36
      \mathchardef\geqslant="3\AMSa3E

      \let\leq=\leqslant \let\le=\leqslant
      \let\geq=\geqslant 
    \fi
  \fi
\fi % End of NFSS release 1

\ifnfsstwo
  \DeclareMathAlphabet{\mathbfit}{OT1}{cmr}{bx}{it}
  \SetMathAlphabet\mathbfit{bold}{OT1}{cmr}{bx}{it}
  \DeclareMathAlphabet{\mathbfss}{OT1}{cmss}{bx}{n}
  \SetMathAlphabet\mathbfss{bold}{OT1}{cmss}{bx}{n}
  \ifAMStwofonts
    \ifCUPmtlplainloaded \else
      \DeclareSymbolFont{UPM}{U}{eur}{m}{n}
      \SetSymbolFont{UPM}{bold}{U}{eur}{b}{n}
      \DeclareSymbolFont{AMSa}{U}{msa}{m}{n}
      \DeclareMathSymbol{\upi}{0}{UPM}{"19}
      \DeclareMathSymbol{\umu}{0}{UPM}{"16}
      \DeclareMathSymbol{\upartial}{0}{UPM}{"40}
      \DeclareMathSymbol{\leqslant}{3}{AMSa}{"36}
      \DeclareMathSymbol{\geqslant}{3}{AMSa}{"3E}

      \let\leq=\leqslant \let\le=\leqslant
      \let\geq=\geqslant 
    \fi
  \fi
\fi % End of NFSS release 2

\ifCUPmtlplainloaded \else
  \ifAMStwofonts \else % If no AMS fonts
    \def\upi{\pi}
    \def\umu{\mu}
    \def\upartial{\partial}
  \fi
\fi

\title[Gravitational waves from cosmological compact binaries]
{Gravitational waves from cosmological compact binaries}
\author[R. Schneider, V. Ferrari, S. Matarrese and S. F. Portegies Zwart]
       {R. Schneider$^1$, V. Ferrari$^1$, S. Matarrese$^{2,3}$ and 
S. F. Portegies Zwart$^{4,5}$\\
        $^{1}$Dipartimento di Fisica ``G. Marconi",
Universit\'a degli Studi di Roma, ``La Sapienza"
and Sezione INFN  ROMA1,\\ piazzale Aldo  Moro
5, 00185 Roma, Italy\\
        $^2$Dipartimento di Fisica ``Galileo Galilei ",
Universit\'a degli Studi di Padova
and Sezione INFN  PADOVA,\\ via Marzolo 8, 35131 Padova, Italy \\
$^3$Max-Planck-Institut f\"ur Astrophysik,  
Karl-Schwarzschild-Strasse 1, D-85748 Garching - Germany \\
$^4$Institute for Astrophysical research,
		  Boston University,
		  725 Commonwealth Ave.,
		  Boston, MA 02215, USA\\
$^5$ Hubble Fellow}

\date{January 2000}

\pagerange{\pageref{firstpage}--\pageref{lastpage}}
\pubyear{2000}
\begin{document}

\maketitle
\label{firstpage}

\begin{abstract}
We consider gravitational waves emitted by various populations of 
compact binaries at cosmological distances. We use population 
synthesis models to characterize the properties of double neutron stars,
double black holes and double white dwarf binaries as well as white 
dwarf-neutron star, white dwarf-black hole and black hole-neutron star
systems. \\
We use the observationally determined cosmic star formation history 
to reconstruct the redshift distribution of these sources 
and their merging rate evolution.\\
The gravitational signals emitted by each source during its early-inspiral
phase add randomly to produce a stochastic background in the low frequency
band with spectral strain amplitude 
between $\sim 10^{-18} \, \mbox{Hz}^{-1/2}$ and
$\sim 5 \times 10^{-17}\,\mbox{Hz}^{-1/2}$ at frequencies in the interval
$[\sim 5 \times 10^{-6}-5 \times 10^{-5}]$~Hz.

The overall signal which, at frequencies above $10^{-4}$~Hz, 
is largely dominated by double white dwarf 
systems, might be detectable with LISA in the frequency range $[1-10]$~mHz
and acts like
a confusion limited noise component which might limit the LISA sensitivity
at frequencies above 1~mHz.
\end{abstract}

\begin{keywords}
gravitation -- stars: formation -- stars: binaries--gravitational waves.
\end{keywords}

\section{Introduction}

Binaries with two compact stars are the most promising sources for
gravitational radiation. The final phase of spiral in may be detected
with ground-based (LIGO, VIRGO, GEO and TAMA) and space-borne laser
interferometers (LISA). 
This has motivated researchers to model
gravitational waveforms and to develop population synthesis codes to
estimate the properties and formation rates of possible sources for
gravitational wave radiation.
 
Since there is not yet a single prescription for calculating the
gravitational emission from a compact binary system, it is customary
to divide the gravitational waveforms in two main pieces: the inspiral
waveform, emitted before tidal distortions become noticeable, and the
coalescence waveform, emitted at higher frequencies during the epoch
of distortion, tidal disruption and/or merger (Cutler \etal 1993). 

As the binary, driven by gravitational radiation reaction, spirals in,
the frequency of the emitted wave increases until the last 3 orbital cycles
prior to complete merger.

With post-Newtonian expansions of the equations of motion for two
point masses, the waveforms can be computed fairly accurately in the
relatively long phase of spiral in (see, for a recent review, Rasio \&
Shapiro 2000 and references therein).
Conversely, the gravitational waveform from the coalescence can only 
be obtained from extensive numerical calculations
with a fully general relativistic treatment. Such calculations 
are now well underway (Brady, Creighton \& Thorne 1998; Damour, Iyer \& 
Sathyaprakash 1998; Rasio \& Shapiro 1999).  
 
In this paper, we consider the low frequency signal from the early phase of the
spiral in, which is of interest for space-borne antennas, such as LISA.
For this purpose, we use the leading order expression for the single
source emission spectrum, obtained using the quadrupole approximation.
Our analysis includes various populations of compact binary systems:
black hole-black hole (bh, bh), neutron star-neutron star (ns, ns),
white dwarf-white dwarf (wd, wd) as well as mixed systems such as
(ns, wd), (bh, wd) and (bh, ns).

For some of these sources [(ns, ns), (wd, wd) and (ns, wd)],
statistical information on the event rate can be inferred from
electromagnetic observations.  In particular, there are several
observational estimates of the (ns, ns) merger rate obtained from
statistics of the known population of binary radio pulsars (Narayan, Piran
\& Shemi 1991; Phinney 1991).

A rather large number of close white dwarf binaries have recently been
found (see Maxted \&
Marsh 1999 and Moran 1999). However, it is customary to
constrain the (wd, wd) merger rate from the observed SNIa rate (see
Postnov \& Prokhorov 1998).  Also the population of binaries where a
radio pulsar is accompanied by a massive unseen white dwarf may be
considerably higher than hitherto expected (Portegies Zwart \&
Yungelson 1999).

Since most stars are members of binaries and the formation rate of
potential sources of gravitational waves may be abundant in the Galaxy,
the gravitational-wave signal emitted by such binaries might produce a
stochastic background.  This possibility has been explored by various
authors, starting from the earliest work of Mironovskij (1965) and Rosi
\& Zimmermann (1976) until the more recent investigations of Hils,
Bender \& Webbink (1990), Lipunov \etal (1995), Bender \& Hils
(1997), Giazotto, Bonazzola \& Gourgoulhon (1997),
Giampieri (1997), Postnov \& Prokhorov (1998),  and Nelemans,
Portegies Zwart \&
Verbunt (1999).  This background, which acts like a noise component
for the interferometric detectors, has always been viewed as a
potential obstacle for the detection of gravitational wave backgrounds
coming from the early Universe.

In this paper we extend the investigation of compact binary systems to
extragalactic distances, accounting for the binaries which have been
formed since the onset of galaxy formation in the Universe.  Following
Ferrari, Matarrese \& Schneider (1999a, 1999b: hereafter referred as FMSI
and FMSII, respectively), we modulate the binary formation rate in the
Universe with the cosmic star formation history derived from
observations of field galaxies out to redshift $z \sim 5$ (see \eg Madau,
Pozzetti \& Dickinson 1998b; Steidel \etal 1999).

The magnitude and frequency distribution of the integrated
gravitational signal produced by the cosmological population of
compact binaries is calculated from the distribution of binary
parameters (masses and types of both stars, orbital separations and
eccentricities). These orbital parameters characterize the 
gravitational-wave signal which we observe on Earth.

Detailed information for the properties of the binary population may
be obtained through population synthesis.  We use the binary
population synthesis code {\sf SeBa} to simulate the characteristics
of the binary population in the Galaxy (Portegies Zwart \& Verbunt
1996; Portegies Zwart \& Yungelson 1998). The characteristics of the
extragalactic population are derived from extrapolating these results
to the local Universe.

The paper is organized as follows: in Section~2 we describe the
population synthesis calculations. Section~3 deals with the energy
spectrum of a single source followed, in Section~4, by the derivation of
the extragalactic backgrounds for the different binary populations. 
In Sections~3 and~4 we also give details on the adopted astrophysical and 
cosmological properties of the systems. 
In Section~5, we compute the spectral strain amplitude
produced by each cosmological population and investigate its
detectability with LISA.  Finally, in Section~6 we
summarize our main results and compare them with other previously
estimated astrophysical background signals.

\section{Population synthesis model}
\label{sec:popSeBa}

In order to characterize the main properties of different compact
binary systems, we use the binary population synthesis program {\sf
SeBa} of Portegies Zwart \& Yungelson (1998). Details of the code can
be found in (Portegies Zwart \& Yungelson 1998). Here, we simply
recall the main assumptions of their adopted model B, which
satisfactorily reproduces the properties of observed high-mass binary
pulsars (with neutron star companions).

The following initial conditions were assumed: the mass of the primary
star $m_1$ is determined using the mass function described by Scalo
(1986) between 0.1 and 100 $\msun$. For a given $m_1$, the mass of the
secondary star $m_2$ is randomly selected from a uniform distribution
between a minimum of 0.1 $\msun$ and the mass of the primary star. The
semi-major axis distribution is taken flat in $\log a$ (Kraicheva
\etal 1978) ranging from Roche-lobe contact up to $10^6$~R$_\odot$
(Abt \& Levy 1978; Duquennoy \& Mayor 1991).  The initial eccentricity
distribution is independent of the other orbital parameters, and is
$\Xi(e) = 2 e$.

Neutron stars receive a velocity kick upon birth.  Following Hartman
\etal (1997), model B assumes the distribution for isotropic kick
velocities (Paczy{\'n}ski 1990),
\begin{equation}
P(u)du = {4\over \pi} \cdot {du\over(1+u^2)^2},
\label{eqkick}\end{equation}
with $u=v/\sigma$ and $\sigma = 600 \,\,\mbox{km}\,\mbox{s}^{-1}$.

The birthrate of the various compact binaries is normalized to the
Type II+Ib/c supernova rate 
(see Portegies Zwart \& Verbunt 1996). The supernova rate of 
0.01 per year was assumed to be constant over the lifetime of
the galactic disc ($\sim 10$~Gyr).

When computing the birth and merger-rates we account for the 
time-delay between the formation of the progenitor system and that 
of the corresponding degenerate binary, $\tau_s$. Its value is 
set by the time it takes for the least massive between the two companion
stars to evolve on the main sequence. For (bh, bh), (ns, ns) and
(bh, ns) systems $\tau_s \lsim 50$~Myr and it is negligible
compared to the assumed lifetime of the galactic disc. Conversely, 
(wd, wd), (ns, wd) and (bh, wd) binaries follow a slower evolutionary clock 
and $\tau_s$ can be considerably larger. The cumulative
probability distribution, $P(<\tau_s)$, predicted  
by the population synthesis code is shown in Fig.~1. For these
systems $\tau_s$ can be as large as 10~Gyr although all systems are predicted
to have $\tau_s\leq 10$~Gyr.
%%%%%%%%%%%%%%%%%%%%%%%%%%%%%%%%%%%%%%%%%%%%%%%%%%%%%%%%%%%%%%%%%%%%%%%%%%%%
\begin{figure}
\begin{center}
\psfig{figure=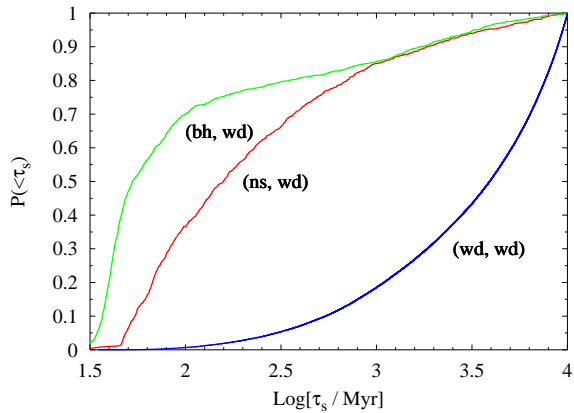,angle=270,width=8cm}
\caption{The cumulative probability distribution for the time delay $\tau_s$
(in Myr) between the formation of the progenitor system and the formation of
the corresponding degenerate binary 
obtained for the (bh, wd), (ns, wd) and (wd, wd) samples.}
\end{center}  
\label{fig:ts}
\end{figure}
%%%%%%%%%%%%%%%%%%%%%%%%%%%%%%%%%%%%%%%%%%%%%%%%%%%%%%%%%%%%%%%%%%%%%%%%%%%%%
 
After the degenerate binary has formed, its further evolution is
determined by the time it takes to radiate away its orbital energy in
gravitational waves. The time between the formation of the degenerate
system and its final coalescence, $\tau_m$, depends on the orbital
configuration and on the mass of the two companion stars.  The
predicted cumulative probability distribution is shown in
Fig.~2 for the (wd, wd), (ns, ns) and (ns, wd) samples.  We
see from the figure that there is a significant fraction of systems
which does not merge in 10~Gyr. For (bh, bh) binaries and mixed
systems with one black hole companion the population synthesis code
predicts very long merger times. In particular, all (bh, bh) systems
appear to have $\tau_m$ greater than 15 Gyr.  The reason for these
large merger times is that binaries with a black hole companion are
characterized by very large initial orbital separations (see \eg
Fig.~3).  In fact, bh progenitors are very massive stars
and have a very strong stellar wind. For this reason they do not
easily reach Roche-lobe over-flow and rarely experience a phase of
mass transfer, which is required to reduce the orbital separation of
the stars. Unfortunately, the evolution (especially the amount of mass
loss in the stellar winds) of high mass stars is rather uncertain
(Langer \etal 1994).  The result that we obtain at least indicates
that it will be very rare to observe any of these bh mergers. Recently
Portegies Zwart \& McMillan (1999) however identified a new channel
for producing black hole binaries which are eligible to mergers on a
relatively short time scale.

%%%%%%%%%%%%%%%%%%%%%%%%%%%%%%%%%%%%%%%%%%%%%%%%%%%%%%%%%%%%%%%%%%%%%%%%%%%%
\begin{figure}
\begin{center}
\psfig{figure=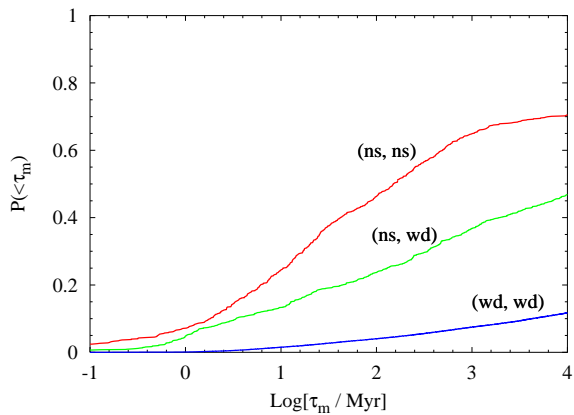,angle=270,width=8cm}
\caption{The cumulative probability distribution for the merger time $\tau_m$
(in Myr) is shown for the (ns, ns), (ns, wd) and (wd, wd) samples.}  
\end{center}
\label{fig:tm}
\end{figure}
%%%%%%%%%%%%%%%%%%%%%%%%%%%%%%%%%%%%%%%%%%%%%%%%%%%%%%%%%%%%%%%%%%%%%%%%%%%%%
%%%%%%%%%%%%%%%%%%%%%%%%%%%%%%%%%%%%%%%%%%%%%%%%%%%%%%%%%%%%%%%%%%%%%%%%%%%%
\begin{figure}
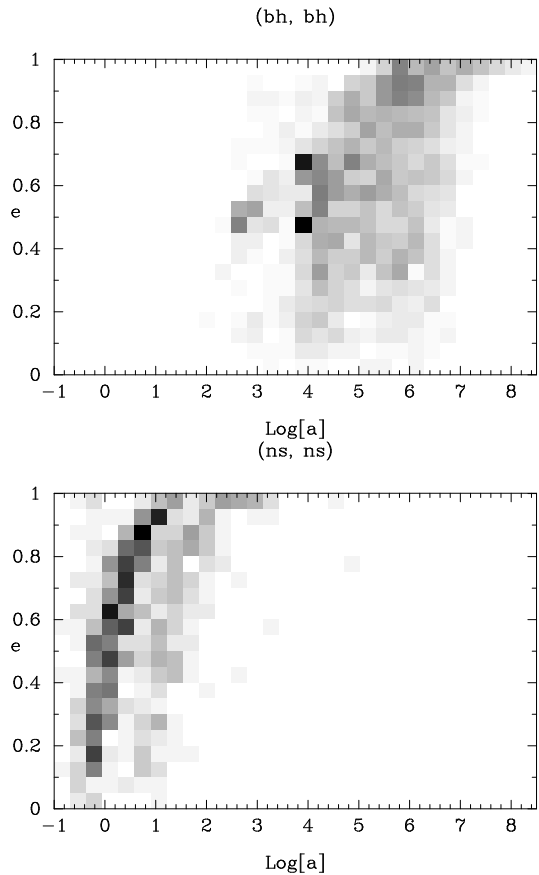

\psfig{figure=bhbhorbit.ps,angle=270,width=7.cm}
\psfig{figure=nsnsorbit.ps,angle=270,width=7.cm}
\caption{The probability distribution for the orbital parameters
of (bh, bh) systems ({\bf left} panel) is compared to that obtained for
the (ns, ns) ({\bf right} panel) population.}  
\label{fig:orbit}
\end{figure}
%%%%%%%%%%%%%%%%%%%%%%%%%%%%%%%%%%%%%%%%%%%%%%%%%%%%%%%%%%%%%%%%%%%%%%%%%%%%%

In Table~\ref{galacticrates}, we summarize the results for all 
binary types that we have investigated.

%%%%%%%%%%%%%%%%%%%%%%%%%%%%%%%%%%%%%%%%%%%%%%%%%%%%%%%%%%%%%%%%%%%%%%%%%%%%
\begin{table}
\caption{The galactic birthrates, $R_{X,gal}$, and merger rates, 
$R^{mrg}_{X,gal}$, obtained for
each compact binary type $X$  using model B of (Portegies Zwart \& 
Yungelson 1998), see text. The rates are normalized to the core-collapse
supernova rate  of $0.01 \,\,\mbox{yr}^{-1}$ and 100\% binarity.
Merger rates are computed after 10~Gyr of the evolution of the Galaxy with
a constant supernova rate.}
\begin{tabular}{@{}ccc@{}}
Binary Type $X$ & $R_{X,gal}\mbox{yr}^{-1}$& 
$R^{mrg}_{X,gal}\mbox{yr}^{-1}$ \\\hline
        &     &   \\[1pt]
(bh, bh) & $6.3 \times 10^{-5}$ & NA\\
(bh, ns) & $1.0 \times 10^{-5}$ & $1.0 \times 10^{-6}$ \\ 
(bh, wd) & $4.4 \times 10^{-5}$ &            $10^{-7}$ \\
(ns, ns) & $3.6 \times 10^{-5}$ & $2.5 \times 10^{-5}$ \\
(ns, wd) & $3.6 \times 10^{-4}$ & $1.6 \times 10^{-4}$ \\
(wd, wd) & $4.4 \times 10^{-2}$ & $4.8 \times 10^{-3}$ \\ 
\end{tabular}
\label{galacticrates}
\end{table}
%%%%%%%%%%%%%%%%%%%%%%%%%%%%%%%%%%%%%%%%%%%%%%%%%%%%%%%%%%%%

\section{Inspiral energy spectrum of single sources}
\label{sec:single}

Assuming that the orbit of the binary system has already been circularized by 
gravitational radiation reaction, the inspiral  
spectrum $dE/d{\nu}$ emitted by a single source 
can be obtained using the quadrupole approximation 
(Misner, Thorne \& Wheeler 1995). The resulting expression,
in geometrical units (G=c=1), can be written as,
\be
\frac{dE}{d\nu}=\frac{\pi}{3}\,\frac{{\cal M}^{5/3}}{(\pi \nu)^{1/3}}
\label{singlespectrum}
\ee   
where ${\cal M}=\mu^{3/5}  M^{2/5}$ is the so-called chirp mass,
$M=m_1+m_2$ stands for the total mass and $\mu=m_1\,m_2/M$ is the 
reduced mass.

The frequency $\nu$ at which gravitational waves are emitted is twice the 
orbital frequency and depends on the time left to the merger event through,
\be
(\pi \nu)^{-8/3} = (\pi \nu_{max})^{-8/3} +  
\frac{256}{5}\,  {\cal M}^{5/3} \,(t_c -t)
\label{eq:emissionfreq}
\ee
where $t_c$ is the time of the final coalescence and we terminate the
inspiral spectrum at a frequency $\nu_{max}$. 
When Post-Newtonian expansion terms are included, it is customary to
consider the inspiral spectrum as a good approximation all the way up to
$\nu_{LSCO}$, \ie the frequency of the quadrupole waves 
emitted at the last stable circular orbit (LSCO) (see \eg Flanagan \& Hughes 1998).
However, for the purposes of our study, we neglect  
post-Newtonian terms
and we set the value of $\nu_{max}$ to correspond to the quadrupole
frequency emitted when the orbital separation is roughly 3 times the
separation at the LSCO. 

The value of the orbital separation at the LSCO depends on the mass
ratio of the two stellar components and varies between $5 M-6 M$. The
lower limit is obtained in the test particle approximation ($m_1 \gg
m_2$), whereas the upper value corresponds to the equal-mass case
($m_1 \sim m_2$) (see Kidder, Will \& Wiseman 1993).  If we consider
the equal-mass limit, which is more conservative for constraining the
maximum frequency, $\nu_{LSCO} \sim 0.022/M$ and $\nu_{max} \sim
0.19\,\, \nu_{LSCO}$.

For (wd, wd) binaries and binaries with one wd companion, 
the maximum frequency, \ie the minimum distance
between the two stellar components, is set in order to cut-off the 
Roche-lobe contact stage. In fact, the mass transfer from one component 
to its companion transforms the original detached binary into a semi-detached
binary. This process can be accompanied by the loss of angular momentum with 
mass loss from the system and the above description cannot be applied. 
Thus, for closed white dwarf binaries we require that the 
mimimum orbital separation is given by $r_{wd}(m_1)+r_{wd}(m_2)$ 
where 
\be
r_{wd}(m)=0.012 R_{\odot} \sqrt{\left(\frac{m}{1.44 M_{\odot}}\right)^{-2/3} -
\,\,\left(\frac{m}{1.44 M_{\odot}}\right)^{2/3}} 
\ee
is the approximate formula for the  radius of a white dwarf from 
Nauenberg (1972) (see also Portegies Zwart \& Verbunt 1996).      

Consider now sources at cosmological distances. The locally measured 
average energy flux emitted by a source at redshift $z$ is,
\be
f(\nu)= \int \frac{d\Omega}{4 \pi}\, \frac{dE}{d\Sigma d\nu}(\nu) =
\frac{(1+z)^2}{4 \pi d_{L}(z)^2} \,\,\frac{c^3}{G} \,\,
\frac{dE_e}{d\nu_e}[\nu (1+z)]
\label{fluxsingle}
\ee
where  $d_{L}(z)$ is the luminosity distance to the source,
$\nu=\nu_e \,(1+z)^{-1}$ is the observed frequency and the factor $c^3/G$
is needed in order to change from geometrical to physical units. 
Thus, the emission spectrum can be written as,
\be
\frac{dE_e}{d\nu_e}[\nu (1+z)]=\frac{\pi}{3}\frac{{\cal M}^{5/3}}{[\pi\,
\nu (1+z)]^{1/3}}  
\label{cosmicspectrum}
\ee
where $\nu$ is the observed frequency emitted by a system at time $t(z)$
\beq
\label{freinterval}
(\pi \nu)^{-8/3}=(\pi \nu_{max})^{-8/3}(1+z)^{8/3} + 
\frac{256}{5} \, {\cal M }^{5/3} \, \\ 
\left[t(z_f)+ \tau_m -t(z)\right](1+z)^{8/3}  \nonumber
\eeq
and we have written the time of the final coalescence $t_c=t(z_c)$ in terms 
of the time of formation $t(z_f)$ and of the merger-time 
$\tau_m = t(z_c) - t(z_f)$.

\section{Extragalactic backgrounds from different binary populations} 

Our main purpose is to estimate the stochastic background signal
generated by different populations of compact binary systems at
extragalactic distances. 

These gravitational sources have
been forming since the onset of galaxy formation in the Universe
and for each binary type $X$ [(ns, ns); (wd, wd); (bh, bh); (ns, wd); 
(bh, wd); (bh, ns)]
we should think of a large ensemble of unresolved and uncorrelated 
elements, each characterized by its masses $m_1$ and $m_2$ (or $M$ and $\mu$),
by its redshift and by its time-delays $\tau_s$ and $\tau_m$ 
[see eqns~(\ref{cosmicspectrum}) and~(\ref{freinterval})].      

Thus, in order to consider all contributions from different elements of 
the ensemble $X$, we must integrate the single source emission spectrum over 
the distribution functions for the masses $M$ and $\mu$, for the time-delays 
and for the redshifts.

The distribution functions for $\tau_s$ and $\tau_m$ depend on the
binary type $X$ and have been derived from the population synthesis code
discussed in the previous section. 
The distribution function for ${\cal M}$ can be similarly estimated. 

However, $\tau_s$, $\tau_m$ and ${\cal M}$ are not 
independent random variables.
In particular, $\tau_m$ and ${\cal M}$ are correlated because ${\cal M}$ 
defines the rate of orbital decay, once the degenerate system has formed.
Thus, for each binary population $X$, we consider the joint probability
distribution for $\tau_s$, $\tau_m$ and ${\cal M}$,
\be
p_X(\tau_s, \tau_m, {\cal M})\,\, d \tau_m \,\,d\tau_s \,\,d{\cal M}.
\ee
 
Conversely, the redshift distribution function, \ie the evolution of the
formation rate for each binary type $X$, can be deduced from the 
observed cosmic star formation history out to $z \sim 5$. 

In the following subsections, we illustrate the procedure we have
followed to derive the birth and merger-rates for all binary populations
and to compute the spectra of the corresponding stochastic gravitational 
backgrounds.  
   
\subsection{Cosmic star formation rate}

\begin{figure}
\psfig{file=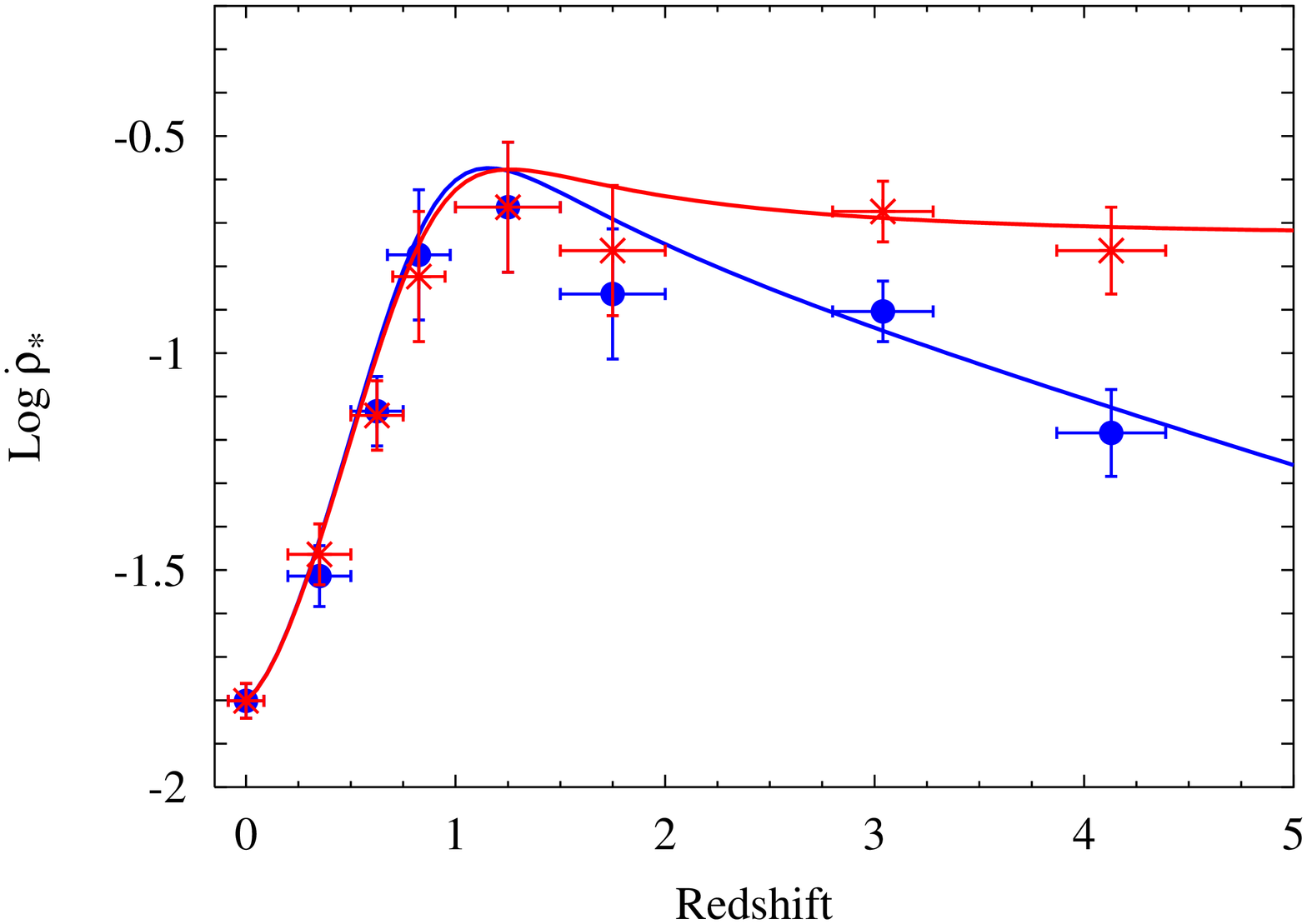,angle=270,width=7.cm}
\psfig{file=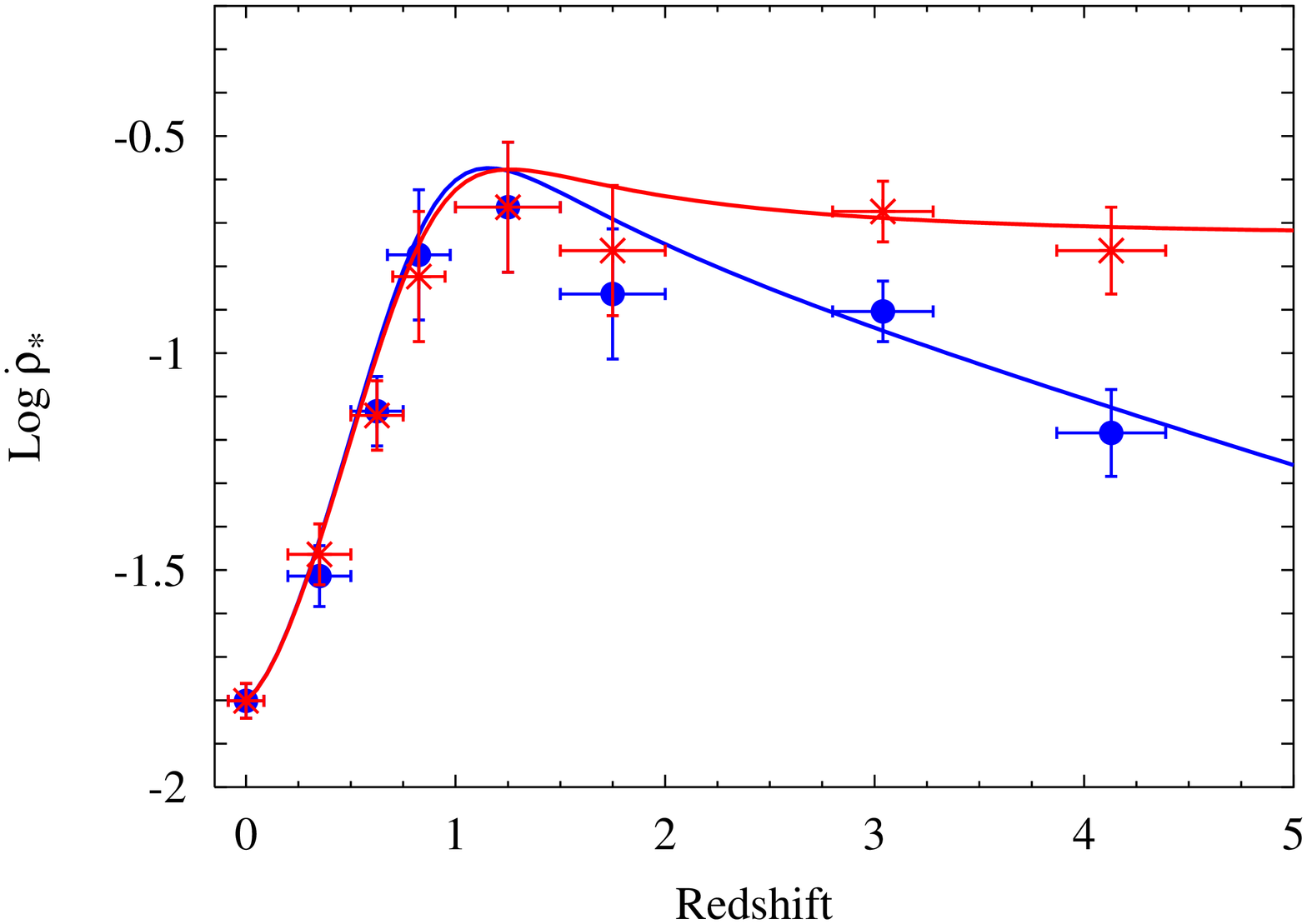,angle=270,width=7.cm}
\caption{The Log of the star formation rate density in units of 
$\msun \mbox{yr}^{-1} \mbox{Mpc}^{-3}$ as a function of redshift for a 
cosmological background model with  $\Omega_{M}=1$, $\Omega_{\Lambda}=0$,
$H_0=50 \,\mbox{km}\mbox{s}^{-1} \mbox{Mpc}^{-1}$ and a 
Scalo (1986) IMF. {\bf Left}: The data points correspond to 
UV, H$\alpha$ and IR observations of field galaxies. ({\bf Filled dots}) 
UV observations  of Treyer \etal (1998), Lilly \etal (1996), 
Connolly \etal (1997), HDF Madau \etal (1996), Steidel \etal (1996),(1999); 
({\bf asterisks}) H$\alpha$ observations of Gallego \etal 1995, Gronwall
1999, Tresse \& Maddox 1998, Glazebrook  \etal 1998, Yan \etal 1999); ({\bf triangles}): ISO IR observations (Flores \etal 1999) and the lower limit
of SCUBA data (Hughes \etal 1998). ({\bf Right}):
The dust-corrected SFR density as derived from UV
data in the two models most favoured by observations predicted by 
Calzetti \& Heckman ({\bf asterisks}) and by Pei, Fall \&
Hauser {\bf filled dots} (see text).} 
\label{sfr}
\end{figure}

In the last few years, the extraordinary advances attained in observational
cosmology have led to the possibility of identifying actively 
star forming galaxies at increasing cosmological look-back times 
(see, for a thorough review, Ellis 1997).
Using the rest-frame UV-optical luminosity as an indicator of the star
formation activity and integrating on the overall galaxy population,
the data obtained with the {\it Hubble Space Telscope} (HST Madau \etal 1996,
Connolly \etal 1997, Madau \etal 1998a) Keck and
other large telescopes (Steidel \etal 1996, 1999) 
together with the 
completion of several large redshift surveys 
(Lilly \etal 1996, Gallego \etal 1995, Treyer \etal 1998)
have enabled, for the first time, to derive coherent models for
the star formation rate evolution throughout the Universe. \\
A collection of some data obtained at different redshifts 
is shown in the left panel of Figure~\ref{sfr} for a flat cosmological
background model with $\Omega_{\Lambda}=0$, $h=0.5$ and a Scalo (1986) IMF 
with masses in the range $0.1-100 \msun$. 
%%%%%%%%%%%%%%%%%%%%%%%%%%%%%%%%%%%%%%%%%%%%
%%%%%%%%%%%%%%%%%%%%%%%%%%%%%%%%%%%%%%%%%%%%
Although the strong luminosity 
evolution
observed between redshift 0 and 1-2 is believed to be quite firmly established,
the behaviour of the
star formation rate at high redshift is still relatively uncertain.
In particular, the decline of the star formation rate density implied by
the $ \lanlge z \ranlge\, \sim 4$ point of the {\it Hubble Deep Field} 
(HDF, see Fig.~\ref{sfr}) 
is now
contradicted by the star formation rate density derived from a new 
ground-based sample of Lyman break galaxies with $\langle z \rangle\,=4.13$ 
(Steidel \etal 1999) which, instead, seems to indicate that the star
formation rate density remains substantially constant at $z>1-2$.
It has been suggested that this discrepancy might be caused by problems
of sample variance in the HDF point at $\lanlge z \ranlge\,=4$ 
(Steidel \etal 1999). \\

Because dust extinction can lead to
an underestimate of the real UV-optical emission and, ultimately, of the 
real star formation activity, the data shown in the left panel 
of Fig.~\ref{sfr} need to be corrected
upwards according to specific models for the evolution of dust opacity
with redshift. In the right panel of Fig.~\ref{sfr}, the data have been
dust-corrected according to factors obtained by Calzetti \& Heckman (1999)
and by Pei, Fall \& Hauser (1999). Using different approaches, these authors 
have recently investigated the cosmic histories of stars, gas, heavy elements 
and dust in galaxies using as inputs the available data from quasar
absorption-line surveys, optical and UV imaging of field galaxies, 
redshift surveys and the COBE DIRBE and FIRAS measurements of the cosmic IR 
background radiation. 
The solutions they obtain appear to reproduce remarkably well a variety of 
observations that were not used as inputs, among which the SFR at various 
redshifts from H$\alpha$, mid-IR and submm
observations and the mean abundance of heavy elements at various epochs 
from surveys of damped Lyman-$\alpha$ systems.

As we can see from the right panel of Fig.~\ref{sfr}, spectroscopic and 
photometric surveys in different wavebands point to a consistent picture 
of the low-to-intermediate redshift evolution:
the SFR density rises rapidly as we go from the local value
to a redshift between $\sim 1-2$ and remains roughly flat between 
redshifts $\sim 2-3$. At higher redshifts, two different
evolutionary tracks seem to be consistent with the data: 
the SFR density might remain substantially constant at $z \gsim 2$
(Calzetti \& Heckman 1999) 
or it might decrease again out to a redshift of 
$\sim 4$ (Pei, Fall \& Hauser 1999). Hereafter, we always indicate the
former model as 'monolithic scenario' and the latter as 'hierarchical
scenario' although this choice is only ment to be illustrative. In fact,
preliminary considerations have pointed out that a constant SFR activity
at high redshifts might not be unexpected in hiererachical structure 
formation models (Steidel \etal 1999). 

Thus, we have updated the star formation rate model that we have 
considered in previous analyses (FMSI, FMSII), even though the
gravitational wave backgrounds are more contributed by low-to-intermediate
redshift sources than by distant ones. In addition,
if a larger dust correction factor should be applied at intermediate
redshifts, this would result in a similar amplification of the 
gravitational background spectra.

\subsection{Birth and merger rate evolution}

Following the method we have previously proposed (FMSI, FMSII), 
for each binary type $X$ the birth and merger-rate evolution could be
computed from the observed star formation rate evolution.
However, this procedure proves to be unsatisfactory because it fails
to provide a fully consistent normalization. 
Its main weakness is that, even if we assume 100\% of binarity, \ie that
all stars are in binary systems, the star formation histories that we have
described above are not corrected for the presence of secondary stars.
For the mass distributions that we have considered,
secondary stars are expected to give a significant contribution 
to the observed UV luminosity as they account for $\sim 1/3$ of the 
fraction of mass in stars more massive than $8 \,\msun$. 

In order to circumvent the necessity of extrapolating 
the UV luminosity indication of massive star formation to the full
range of stellar masses predicted by the model, 
we could directly normalize to the rate of core-collapse supernovae.
This is consistent with the adopted normalization for galactic rates.  

The core-collapse supernova rate  can be directly 
derived from the observed UV luminosity at each redshift, as stars
which dominate the UV emission from a galaxy are the same stars which,
at the end of the nuclear burning, explode as Type II+Ib/c SNae. 
Moreover, the supernova rate is observed independently of the SFR. Therefore
it can be used as an alternative normalization.
%%%%%%%%%%%%%%%%%%%%%%%%%%%%%%%%%%%%%%%%%%%%%%%%%%%%%%%%%%%%%%%%%%%%%%%%%%%%
\begin{figure}
\centerline{\psfig{figure=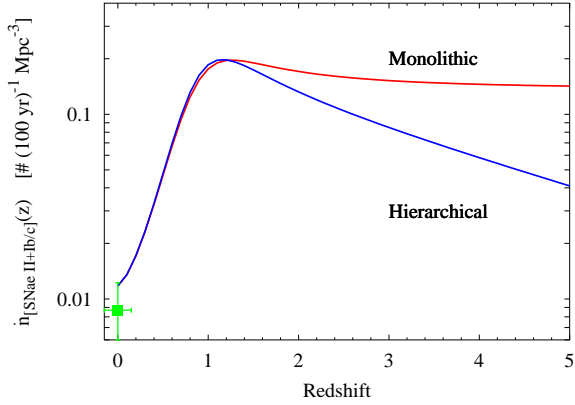,angle=270,width=8cm}}
\caption{The rest-frame frequency of core-collapse SNae  
vs redshift predicted by the monolithic and 
hierarchical models. The predictions are  consistent
with the observed value for the present-day galaxy population (see text).}  
\label{sneII}
\end{figure}
%%%%%%%%%%%%%%%%%%%%%%%%%%%%%%%%%%%%%%%%%%%%%%%%%%%%%%%%%%%%%%%%%%%%%%%%%%%%%

The rates of core-collapse supernovae predicted by the models shown in
Fig.~\ref{sfr} are shown 
in Fig.~\ref{sneII} assuming a 
flat cosmological background model with zero cosmological constant 
and $h=0.5$.

In the same figure, we have plotted the available observations for the 
core-collapse supernova frequency in the local Universe
(Cappellaro \etal 1997, Tamman \etal 1994, Evans \etal 1989, 
see also Madau \etal 1998b).

The binary birthrate per entry per year and comoving
volume $\dot{\eta}(z)$ can be related to the core-collapse supernova rate 
$\dot{n}_{[SNae II+Ib/c]}(z)$ shown in Fig.~\ref{sneII} in the following way,
\be
\dot{\eta}(z)=\frac{\dot{n}_{[SNae II+Ib/c]}(z)}{N_{[SNae II+Ib/c]}}
\label{eq:norm}
\ee
where $N_{[SNae II+Ib/c]}$ is the total number of core-collapse supernovae that
we find in the simulation.

In order to estimate, from $\dot{\eta}(z)$, 
the birth and merger-rate evolution  
of a degenerate binary population $X$, we need to multiply 
eq.~(\ref{eq:norm})
by the number of type $X$ systems in the simulated samples, $N_X$, 
and we also need to properly account for both $\tau_s$ and
$\tau_m$. 

We shall assume that the redshift at the onset of galaxy formation 
in the Universe
is $z_F=5$ and that a zero-age main sequence binary forms at a redshift $z_s$.
After a time interval $\tau_s$, the system has evolved into a  
degenerate binary. Then, 
the redshift of formation of the degenerate binary system, $z_f$, 
is defined as $t(z_f)=t(z_s)+\tau_s$. 
The system then evolves 
according to gravitational wave reaction until, 
after a time interval $\tau_m$, it finally merges.     
Thus, the redshift at which coalescence occurs, $z_c$, is given by 
$t(z_c)=t(z_f)+\tau_m$.

This simple picture implies that the number of $X$ systems formed
per unit time and comoving volume at redshift $z_f$ is
\beq
\dot{n}_{X}(z_f)&=&\int\!\! d\tau_m \, d{\cal M}\int_{0}^{t(z_f)-t(z_F)} \!\!d\tau_s\,\, f_{X} \\ 
& &\frac{\dot{n}_{[SNae II+Ib/c]}(z_s)}{(1+z_s)}
\,p_{X}(\tau_s,\tau_m, {\cal M}) \nn
\eeq
where $f_{X}=N_{X}/N_{[SNae II+Ib/c]}$ and 
$z_s$ is defined by $t(z_s)=t(z_f)-\tau_s$.

If we write, 
\be
\label{eq:deltaprob}
p_{X}(\tau_s,\tau_m, {\cal M})=\frac{1}{N_{X}} \sum_{i}^{N_{X}}
\,\delta(\tau_s -\tau_{s,i})\,\delta(\tau_m - \tau_{m,i})\,\delta({\cal M}-
{\cal M}_i)
\ee
where $\tau_{s,i}$, $\tau_{m,i}$ and ${\cal M}_i$ indicate the time delays 
and the chirp mass for the $i^{th}$ 
element of the ensemble $X$, the birthrate reads,
\beq
\label{eq:birthrate}
\dot{n}_{X}(z_f)&=&\frac{1}{N_{[SNae II+Ib/c]}} \,\,\sum_{i}^{N_{X}}\,
\frac{\dot{n}_{[SNae II+Ib/c]}(z_s)}{(1+z_s)}\\ 
& &\Theta[t(z_f)-t(z_F)-\tau_{s,i}] \nn
\eeq
where $\Theta(x)$ is the step-function.

Similarly, the number of $X$ systems per unit time and comoving volume 
which merge at redshift $z_c$ is,
\beq
\dot{n}^{mrg}_{X}(z_c)=\int_{0}^{t(z_c)-t(z_F)}\!\! d\tau_m 
\int_{0}^{t(z_c)-\tau_m-t(z_F)}\!\! d\tau_s \,\,\, f_{X} \\  
\frac{\dot{n}_{[SNae II+Ib/c]}(z_s)}{(1+z_s)}\int d{\cal M} \,\, 
p_{X}(\tau_s,\tau_m, {\cal M}) \nn
\eeq
where $z_s$ is defined by $t(z_s)=t(z_c)-\tau_m-\tau_s$.  
If we apply eq.~(\ref{eq:deltaprob}), we can write the merger-rate in a form 
similar to eq.~(\ref{eq:birthrate}), \ie
\beq
\label{eq:mergerate}
\dot{n}^{mrg}_{X}(z_c)&=&\frac{1}{N_{[SNae II+Ib/c]}}\,\, \sum_{i}^{N_{X}}
\, \frac{\dot{n}_{[SNae II+Ib/c]}(z_s)}{(1+z_s)}\\ 
& & \Theta[t(z_c)-t(z_F)-\tau_{s,i}-\tau_{m,i}]. \nn
\eeq 

Using this procedure, we compute the birth and merger-rates for all
the synthetic binary populations. The results are presented in 
Figs.~6,~7 and~8.

Due to their relatively small $\tau_s$ 
compared to the
cosmic time, the birthrates of (bh, bh), (ns, ns) and (bh, ns) systems
closely trace the UV-luminosity evolution, although with different
amplitudes. Our simulation suggests that (bh, bh) systems are more 
numerous than (ns, ns) or (bh, ns) (see Fig.~6).
%%%%%%%%%%%%%%%%%%%%%%%%%%%%%%%%%%%%%%%%%%%%%%%%%%%%%%%%%%%%%%%%%%%%%%%%%%%%
\begin{figure}
\begin{center}
\psfig{figure=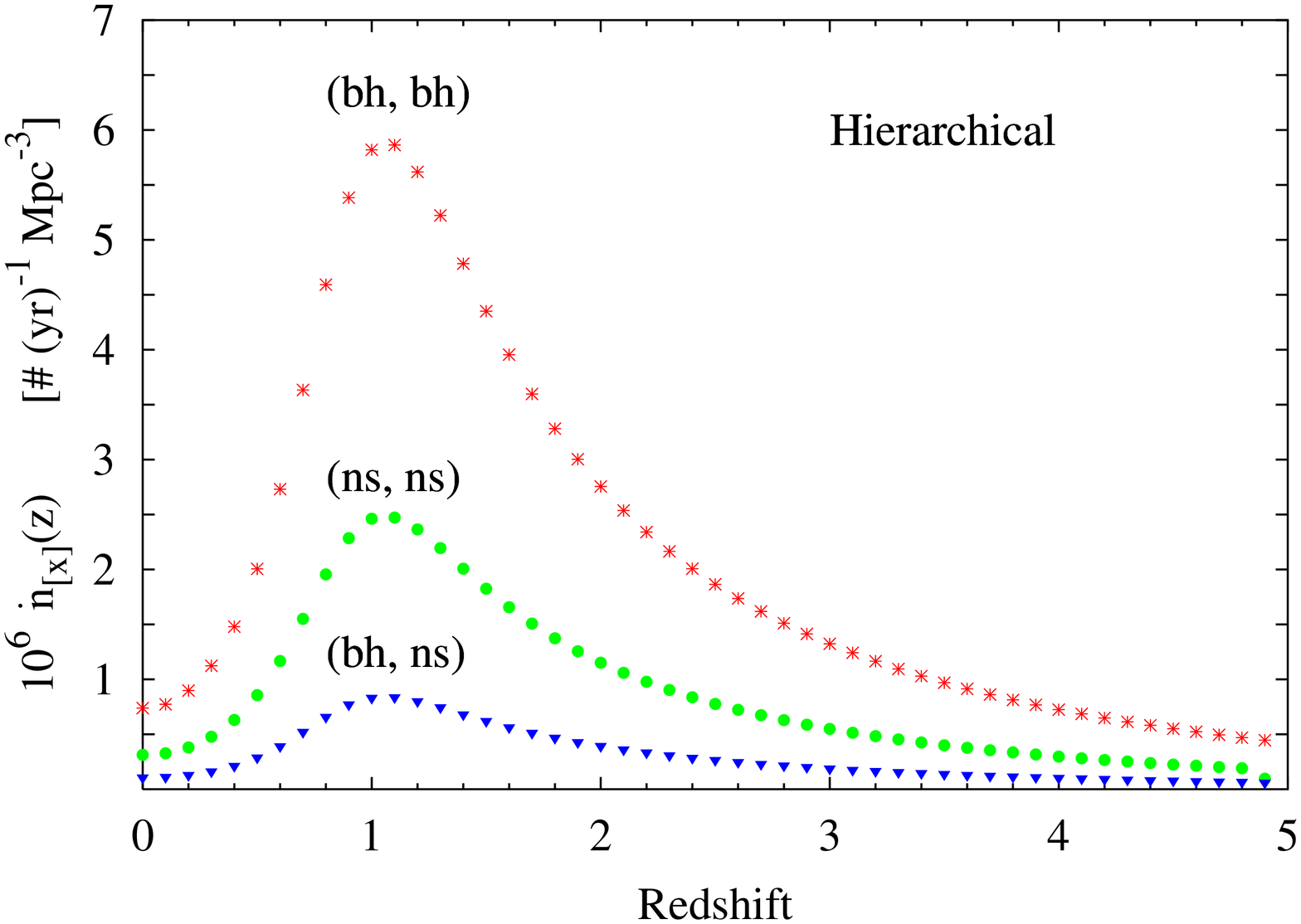,angle=270,width=8cm}
\psfig{figure=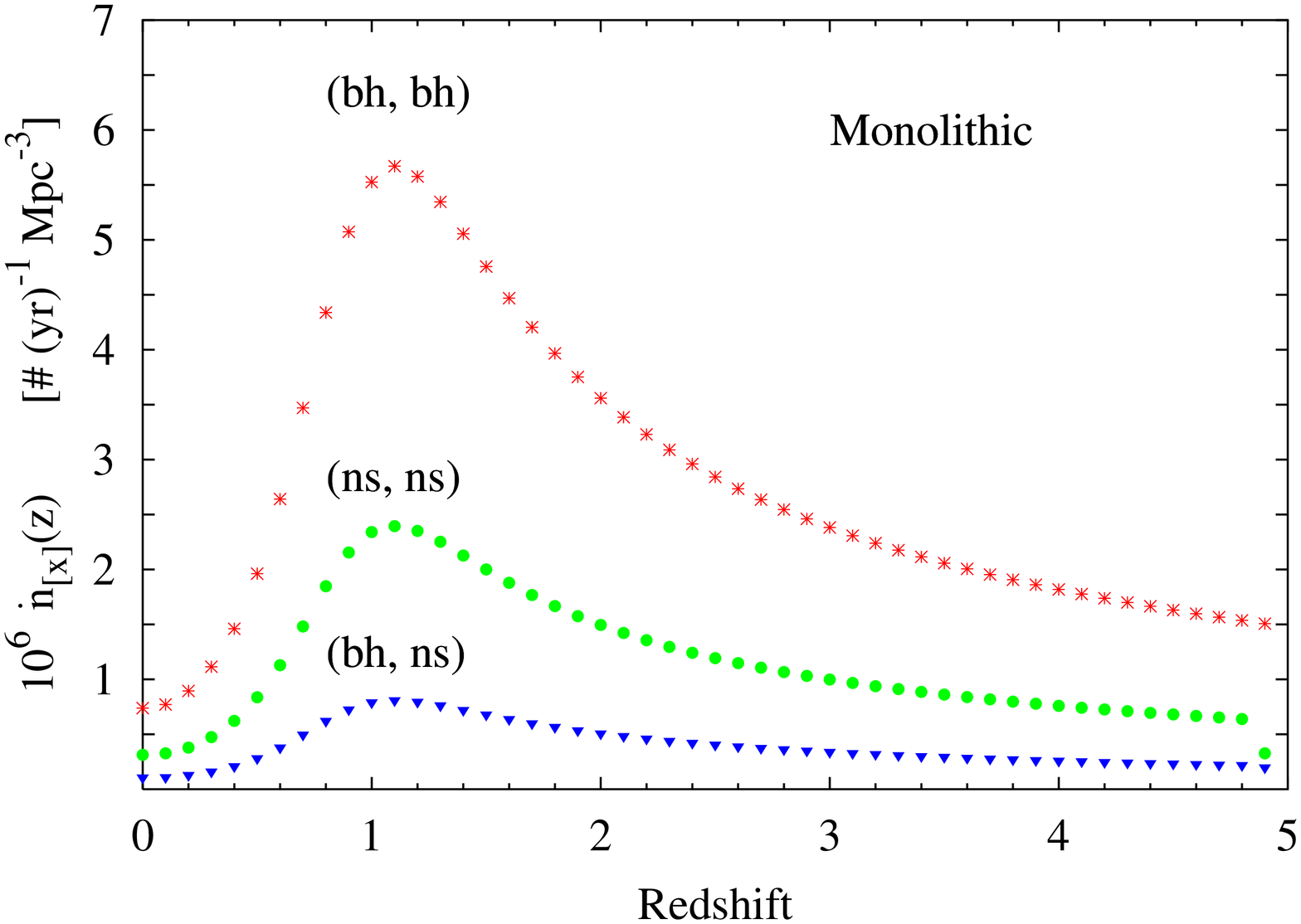,angle=270,width=8cm}
\caption{The formation rate of (bh, bh) ({\bf asterisks}), (ns, ns) ({\bf dots}) and (bh, ns) ({\bf triangles}) binaries  as a function of $z$ is
shown for hierarchical ({\bf upper} panel) and monolithic ({\bf lower} panel) scenarios for 
a flat cosmological background.}  
\end{center}
\label{birthrate1}
\end{figure}
%%%%%%%%%%%%%%%%%%%%%%%%%%%%%%%%%%%%%%%%%%%%%%%%%%%%%%%%%%%%%%%%%%%%%%%%%%%%%

Conversely, Fig.~7 shows that the birthrates of 
(wd, wd), (ns, wd) and (bh, wd) systems misrepresent the original
UV-luminosity evolution as a consequence of their large $\tau_s$. 
The largest is the characteristic time-delay
$\tau_s$, the more the maximum is shifted versus lower redshifts because
the intense star formation activity observed at $z\gsim2$, especially for
monolithic scenarios, boosts
the formation of degenerate systems at $z\lsim2$. 
For hierarchical scenarios,
if the redshift at which significant star formation begins to occur is
$z_F \sim 5$, the birthrate of degenerate systems at redshifts  $\gsim4$
is almost negligible.
%%%%%%%%%%%%%%%%%%%%%%%%%%%%%%%%%%%%%%%%%%%%%%%%%%%%%%%%%%%%%%%%%%%%%%%%%%%%
\begin{figure}
\begin{center}
\psfig{figure=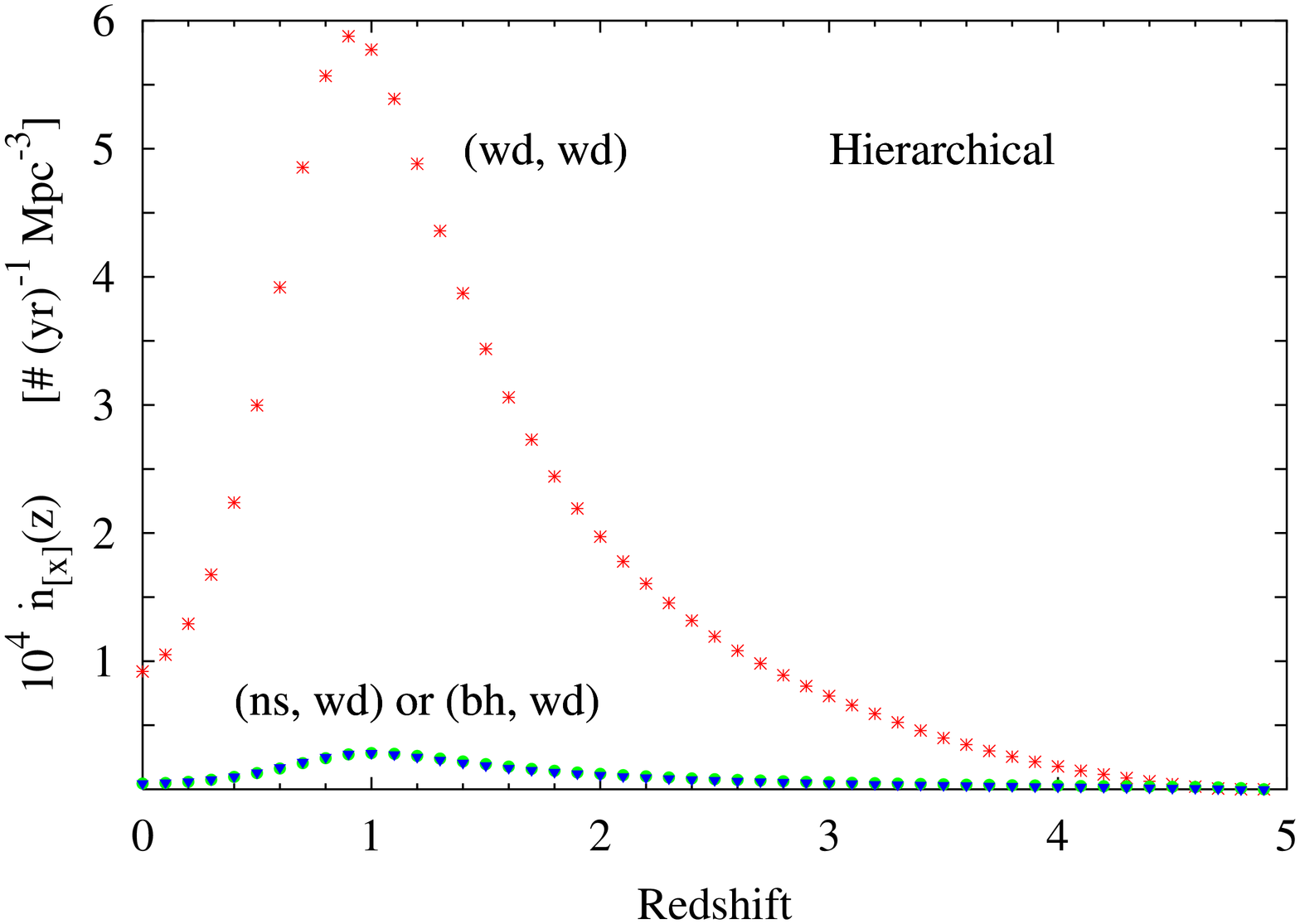,angle=270,width=8cm}
\psfig{figure=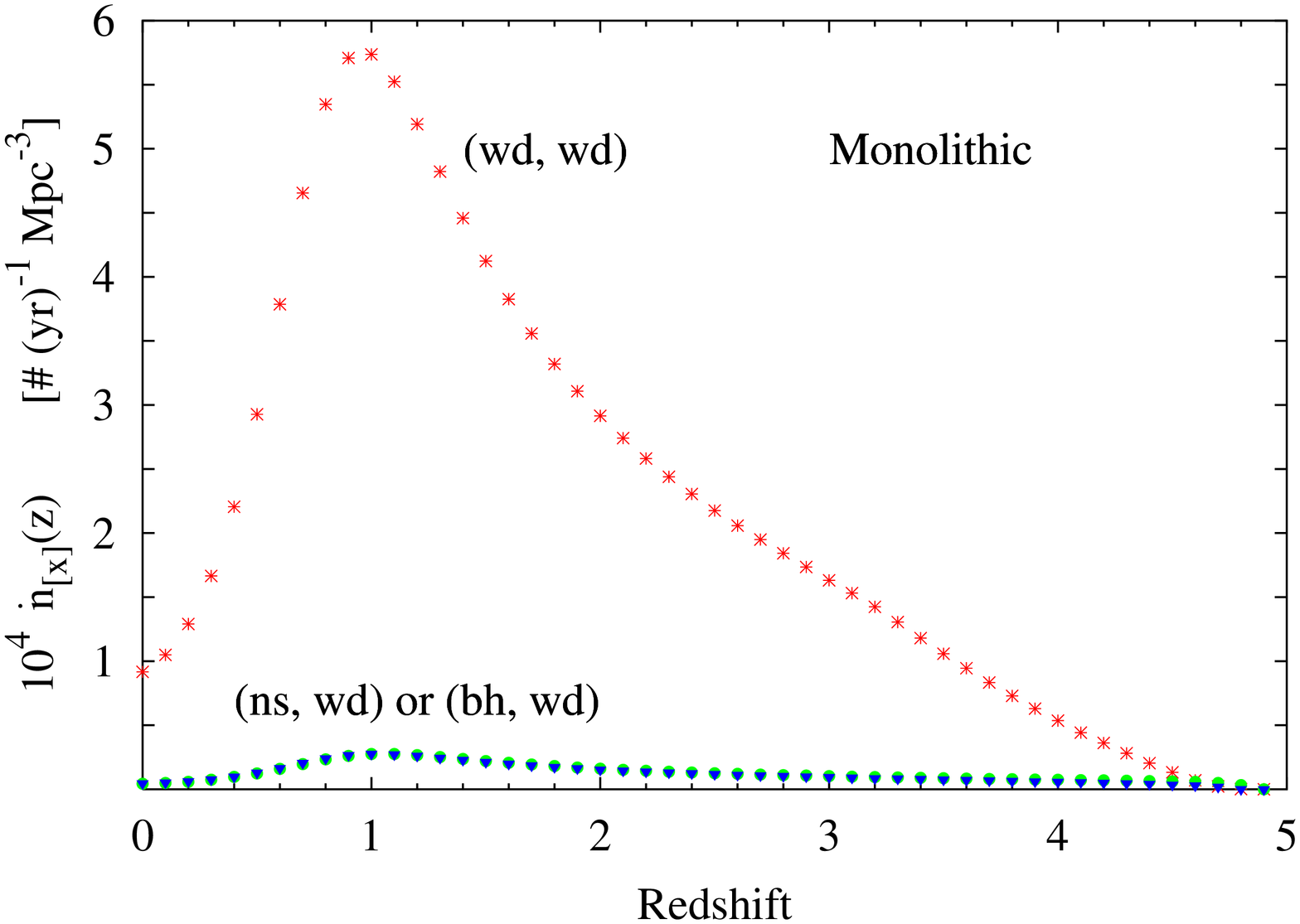,angle=270,width=8cm}
\caption{The formation rate of (wd, wd) ({\bf asterisks}), (ns, wd)
({\bf dots}) and (bh, wd) ({\bf triangles}) binaries  as a function of $z$ is
shown for hierarchical ({\bf upper} panel) and monolithic ({\bf lower} panel) scenarios for 
a flat cosmological background.}
\end{center}  
\label{birthrate2}
\end{figure}
%%%%%%%%%%%%%%%%%%%%%%%%%%%%%%%%%%%%%%%%%%%%%%%%%%%%%%%%%%%%%%%%%%%%%%%%%%%%%

Finally, in Fig.~8,  we have shown the predicted 
merger-rate for (wd, wd), (ns, ns) and (ns, wd) systems. In this case,
the distortion of the original UV-luminosity evolution is even more apparent, 
particularly for monolithic scenarios. The redshift at which the 
maximum merger-rate occurs as well as the high redshift tail reflects
the different $\tau_m$ distributions of these populations. 
We have not shown the merger-rates for 
(bh, bh), (bh, wd) and (bh, ns) binaries because, as we have discussed
in the previous section, these systems are predicted to have
merger-rates consistent with zero throughout the history of the Universe
as a consequence of their very large initial orbital separations.
%%%%%%%%%%%%%%%%%%%%%%%%%%%%%%%%%%%%%%%%%%%%%%%%%%%%%%%%%%%%%%%%%%%%%%%%%%%%
\begin{figure}
\begin{center}
\psfig{figure=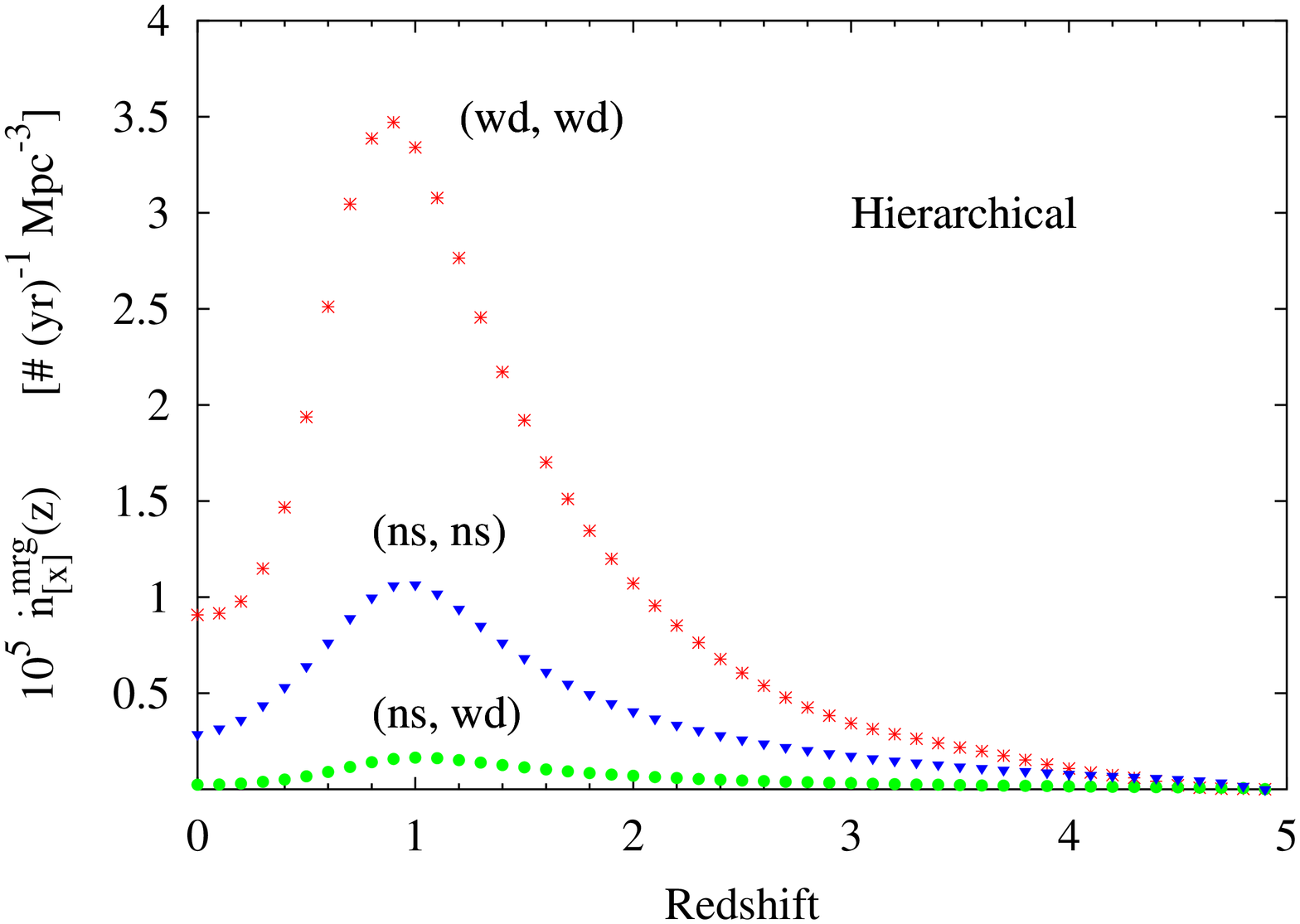,angle=270,width=8cm}
\psfig{figure=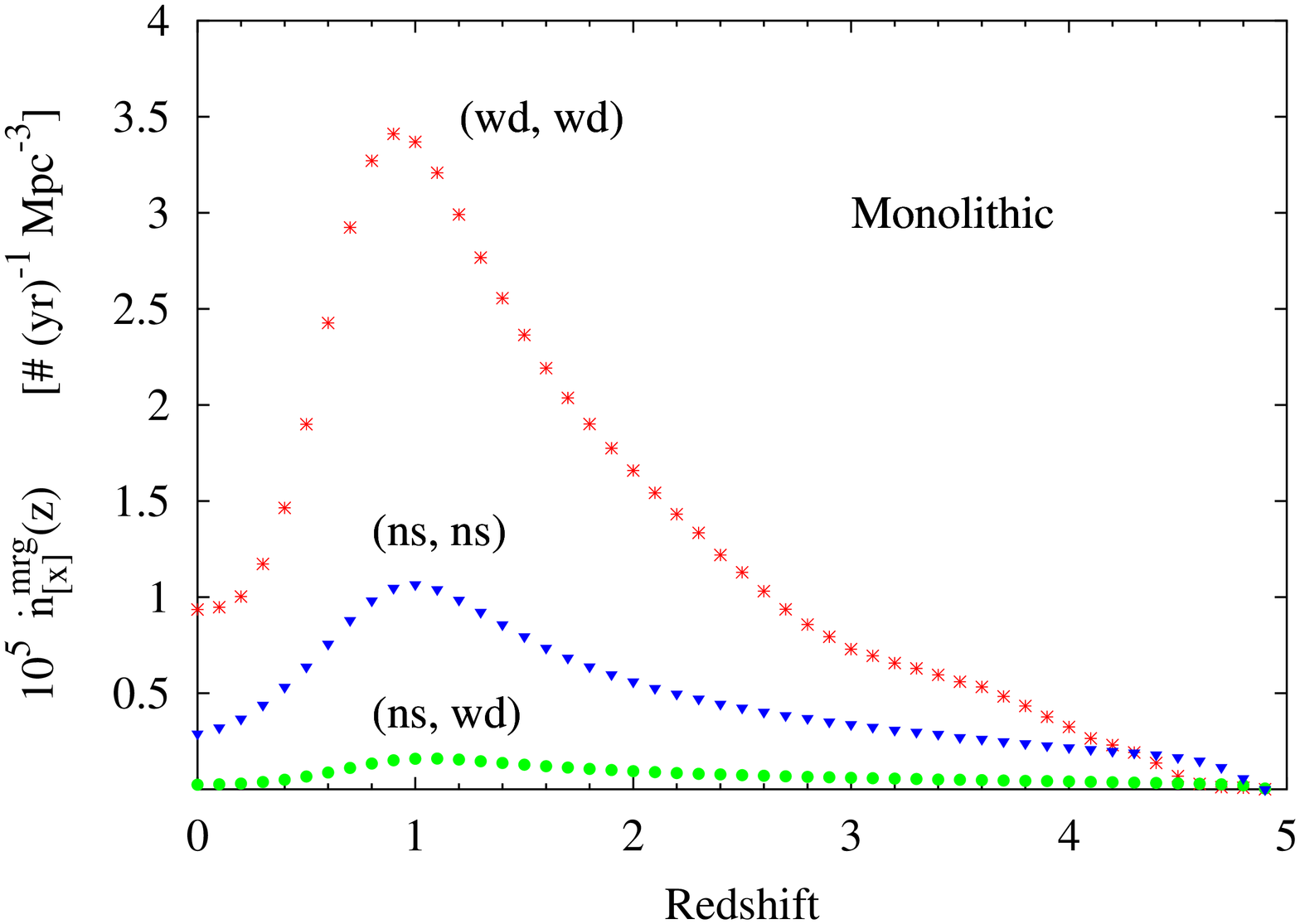,angle=270,width=8cm}
\caption{The merger-rate of (wd, wd) ({\bf asterisks}), (ns, ns) ({\bf
dots}) and (ns, wd) ({\bf triangles}) binaries  as a function of $z$ is
shown for hierarchical ({\bf upper} panel) and monolithic ({\bf lower} panel) scenarios for 
a flat cosmological background.}
\end{center}  
\label{mergerate}
\end{figure}
%%%%%%%%%%%%%%%%%%%%%%%%%%%%%%%%%%%%%%%%%%%%%%%%%%%%%%%%%%%%%%%%%%%%%%%%%%%%%

\subsection{Stochastic backgrounds}

Having characterized each ensemble $X$ by the distribution of chirp mass 
and time delays, $p_X(\tau_s, \tau_m, {\cal M})$, 
and by the birthrate density evolution per entry $\dot{\eta}(z)$, we can sum up
the gravitational signals coming from all the elements of the ensemble. 
The spectrum of the resulting stochastic background, for a binary
type $X$ and at a given observation frequency $\nu$, is
given by the following expression,
 
\beq
\label{eq:BINenergydensity}
\frac{dE}{d\Sigma dt d\nu}[\nu] &=&\int_0^{z_F}\!\!dz_f \int_{0}^{t(z_f)-t(z_F)}
\!\!\!d\tau_s \frac{ N_X \, \dot{\eta}(z_s)}{(1+z_s)} \\ 
& & \int_0^{\infty}\!\!d{\cal M} \,d\tau_m\, 
p_X(\tau_s, \tau_m, {\cal M}) \, \frac{dV}{dz_e^*} \, f[\nu, z_e^*] \nn
\eeq
where $z_F$ is the redshift of the onset of star formation in the Universe, 
$z_f$ is the redshift of formation of the degenerate binary systems,
$z_s$ is the redshift of formation of the corresponding progenitor system
defined by $t(z_s)=t(z_f)-\tau_s$, $f[\nu, z_e^*]$ is given by eq.~(5) and 
$z_e^*$ is the redshift of emission that an element of the ensemble 
must have in order to contribute to the energy density at the observation 
frequency $\nu$.

It follows from eq.~(\ref{freinterval}) that, for a given observation 
frequency $\nu$, $z_e^*$ is a function of $z_f$, $\tau_m$, ${\cal M}$ and $\nu_{max}$.
In principle, an inspiraling compact binary system emits a continuous
signal from its formation to its final coalescence thus,
$z_c \le z_e \le z_f$.
However, in eq.~(\ref{eq:BINenergydensity}) we do not restrict to systems 
which reach their final coalescence at $z_c\geq0$ as we are interested to
{\it any} source between $z=0$ and $z=z_F$ emitting gravitational
waves during its early inspiral phase. Therefore, the signals which contribute
to the local energy density at observation frequency $\nu$ might be emitted
anywhere between $\sup[0,z_c] \le z_e^* \le z_f$, provided that,       
\beq
\label{eq:bin_freconstraint}
(\pi \nu)^{-8/3}=(\pi \nu_{max})^{-8/3}(1+z_e^*)^{8/3} + 
\frac{256}{5} \, {\cal M}^{5/3} \\ 
\left[t(z_f)+ \tau_m -t(z_e^*)\right](1+z_e^*)^{8/3}. \nn  
\eeq
Substituting eq.~(\ref{eq:deltaprob}) in eq.~(\ref{eq:BINenergydensity}),
we can write the background energy density generated by a population $X$ in the
form, 
\beq
\frac{dE}{d\Sigma dt d\nu}[\nu] &=&\int_0^{z_F}\!\! dz_f \sum_{i}^{N_{X}}
\frac{\dot{\eta}(z_s)}{(1+z_s)}\\ 
& &\Theta[t(z_f)-t(z_F)-\tau_{s,i}]\,
\frac{dV}{dz_e^*} \, f[\nu, z_e^*] \nn
\eeq
where $z_e^*$ satisfies eq.~(\ref{eq:bin_freconstraint}).

The predicted spectral energy densities for the populations of degenerate binary types 
that we have considered are plotted in Fig.~9.
For each binary type, we show the results obtained assuming both monolithic
and hierarchical scenarios for the evolution of the underlying galaxy population.

%%%%%%%%%%%%%%%%%%%%%%%%%%%%%%%%%%%%%%%%%%%%%%%%%%%%%%%%%%%%%%%%%%%%%%%%%%%%
\begin{figure*}
\begin{center}
\leavevmode
\psfig{figure=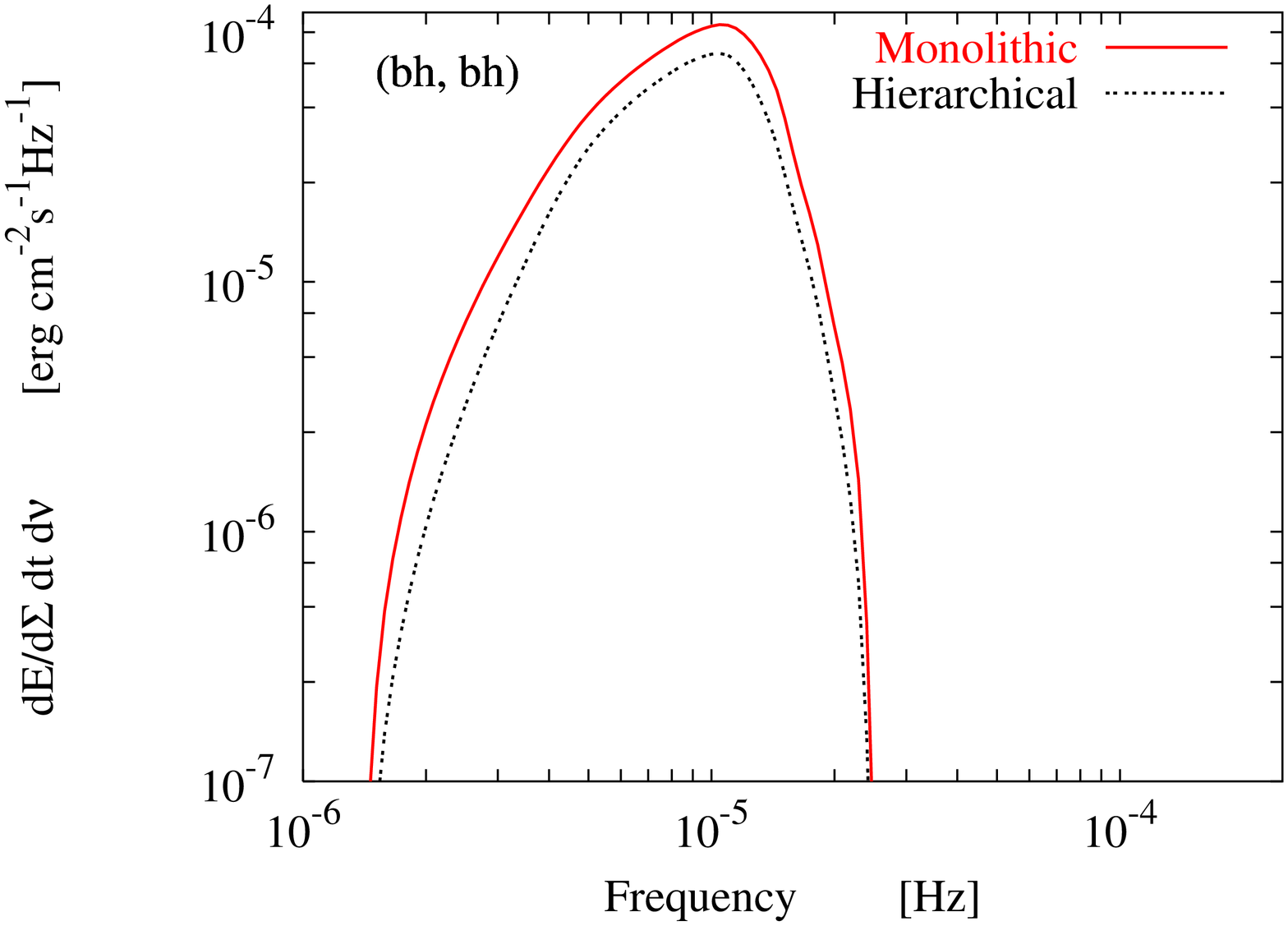,angle=270,width=7.cm}
\psfig{figure=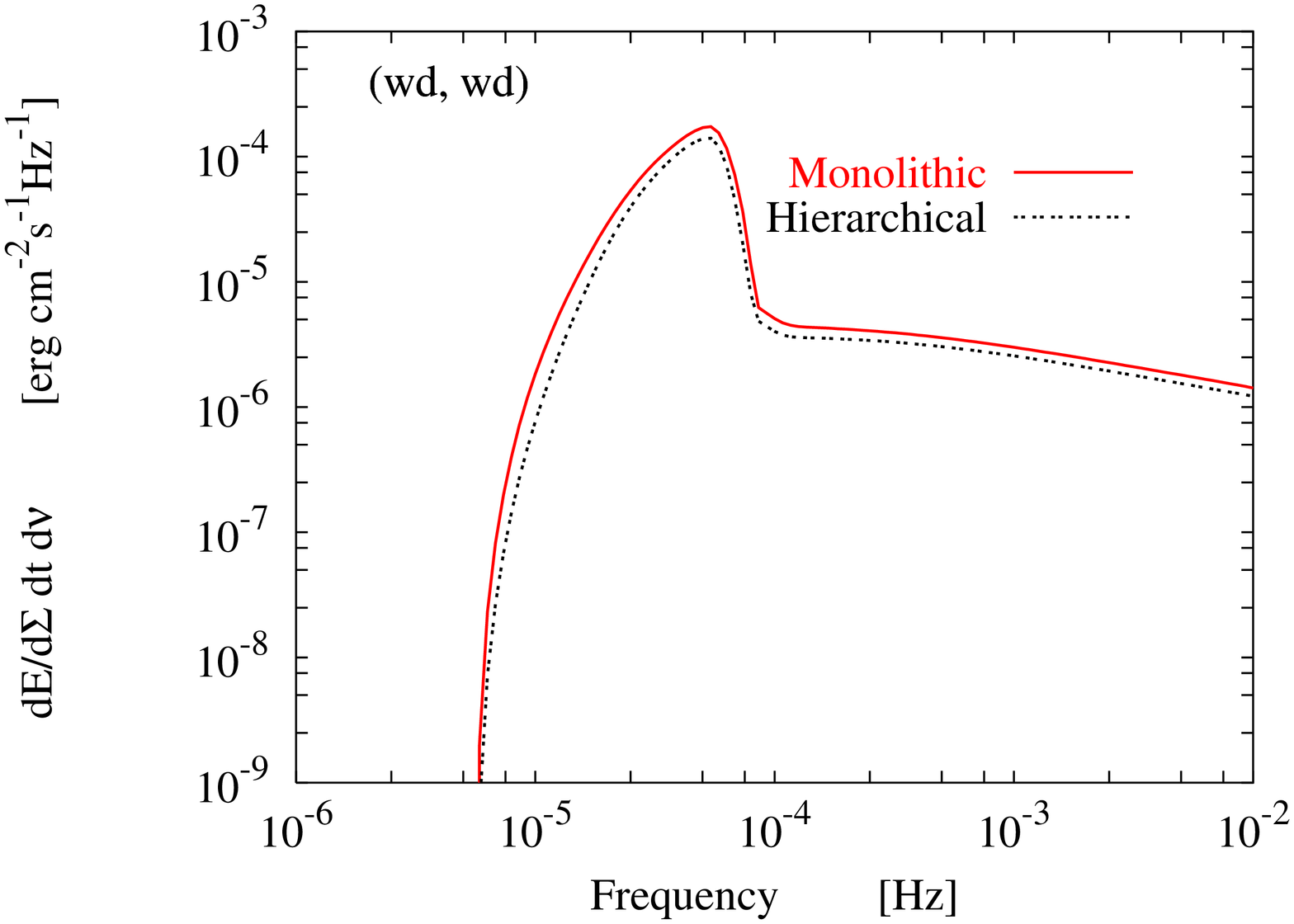,angle=270,width=7.cm}
\psfig{figure=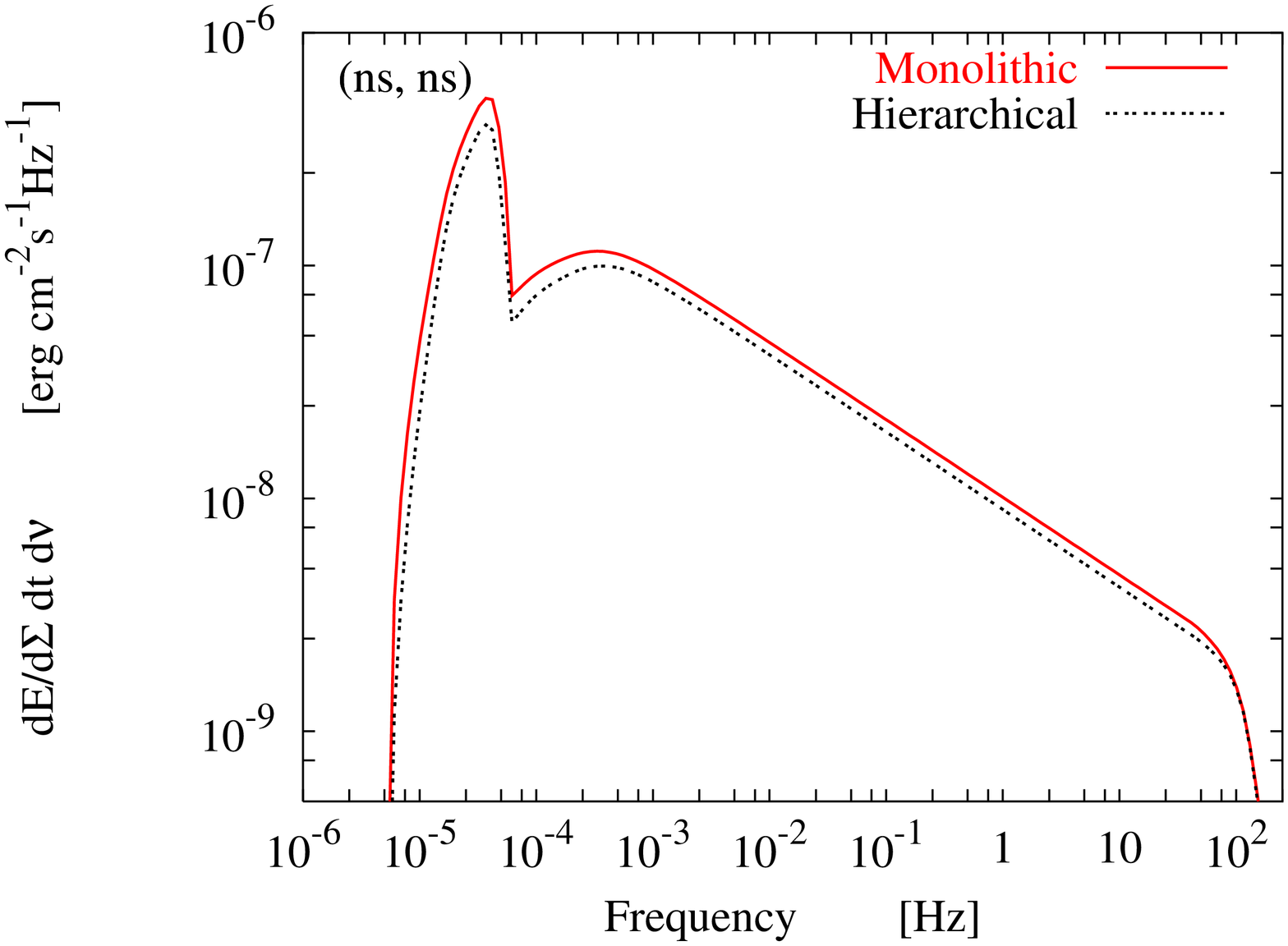,angle=270,width=7.cm}
\psfig{figure=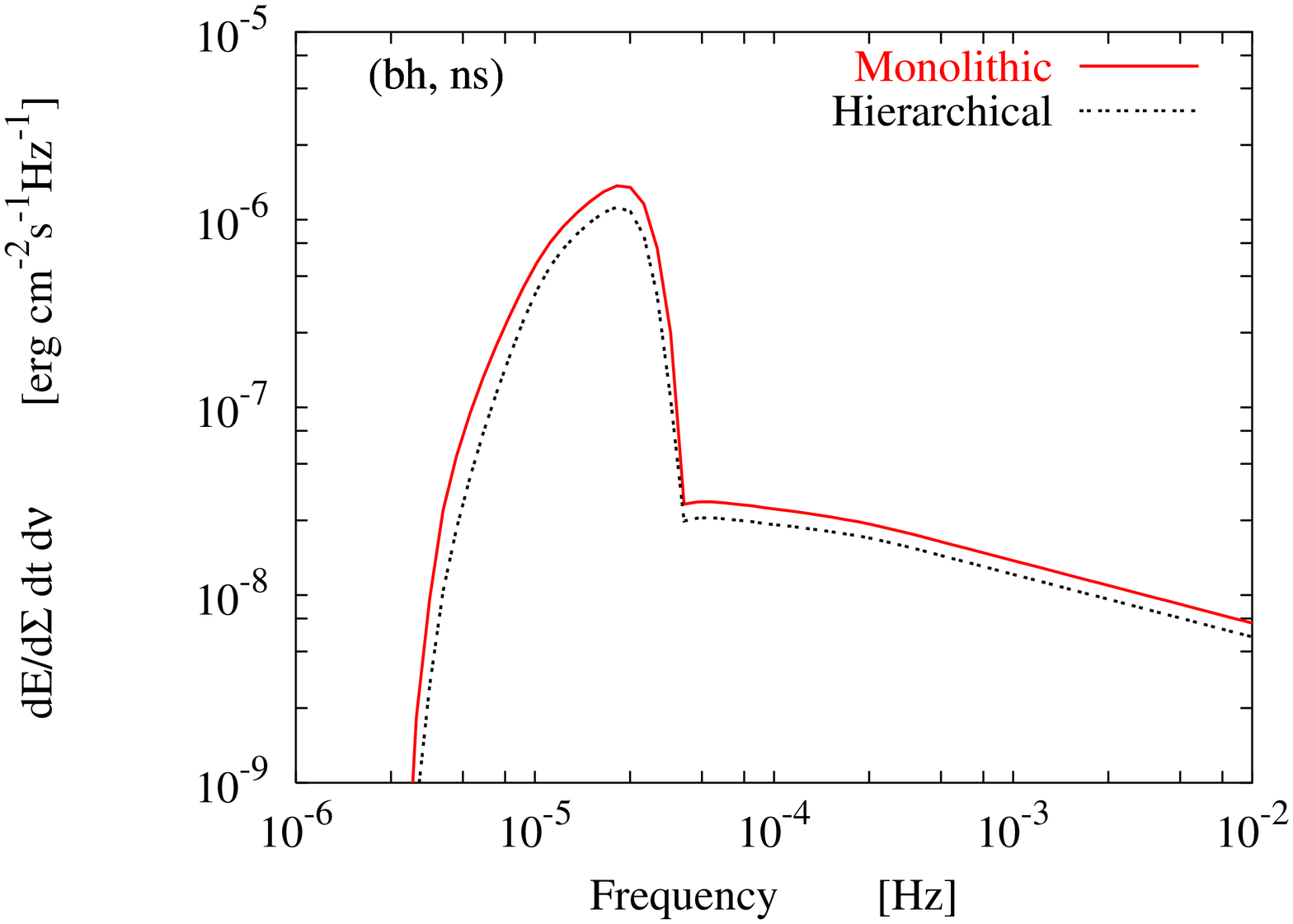,angle=270,width=7.cm}
\psfig{figure=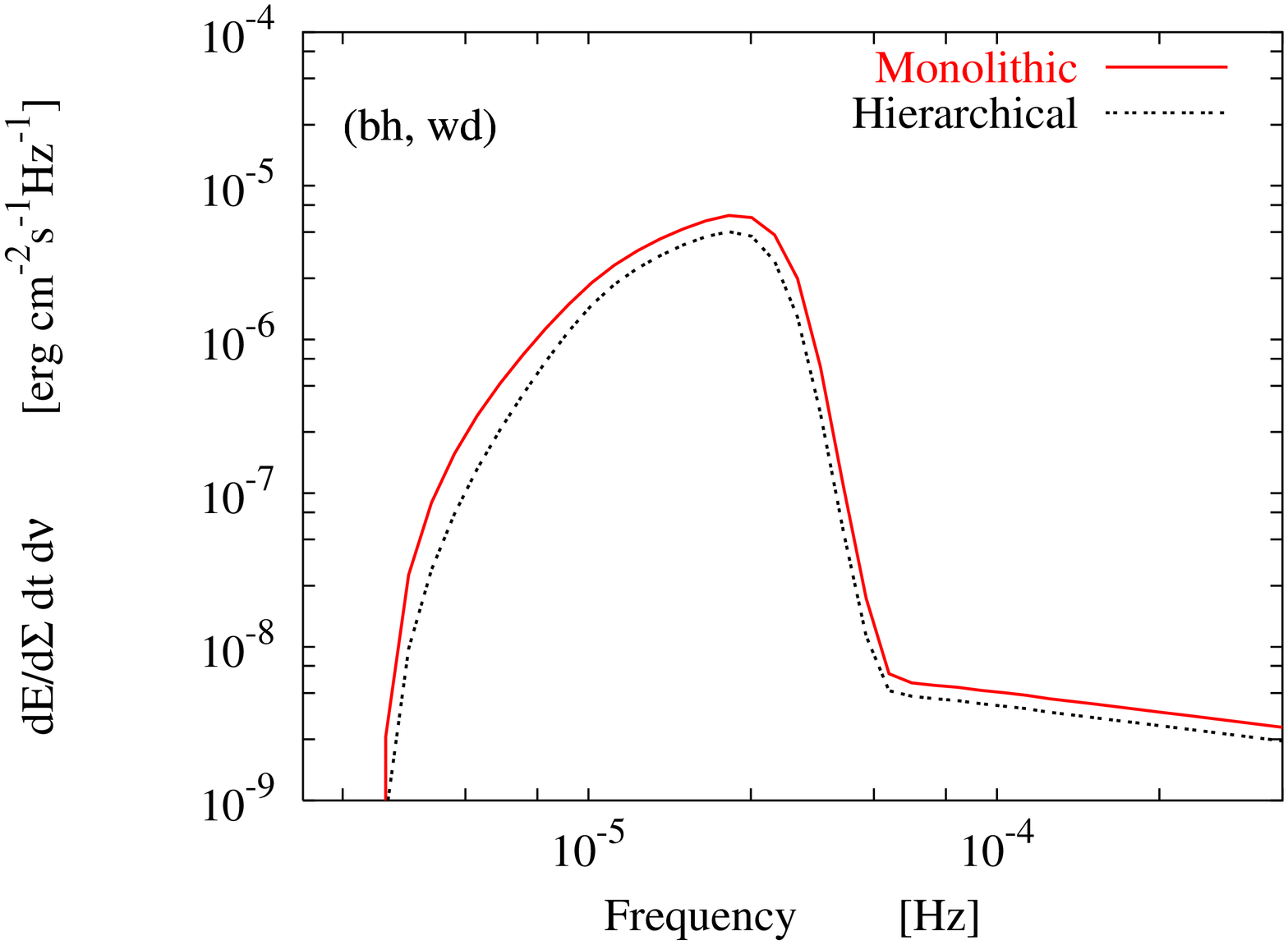,angle=270,width=7.cm}
\psfig{figure=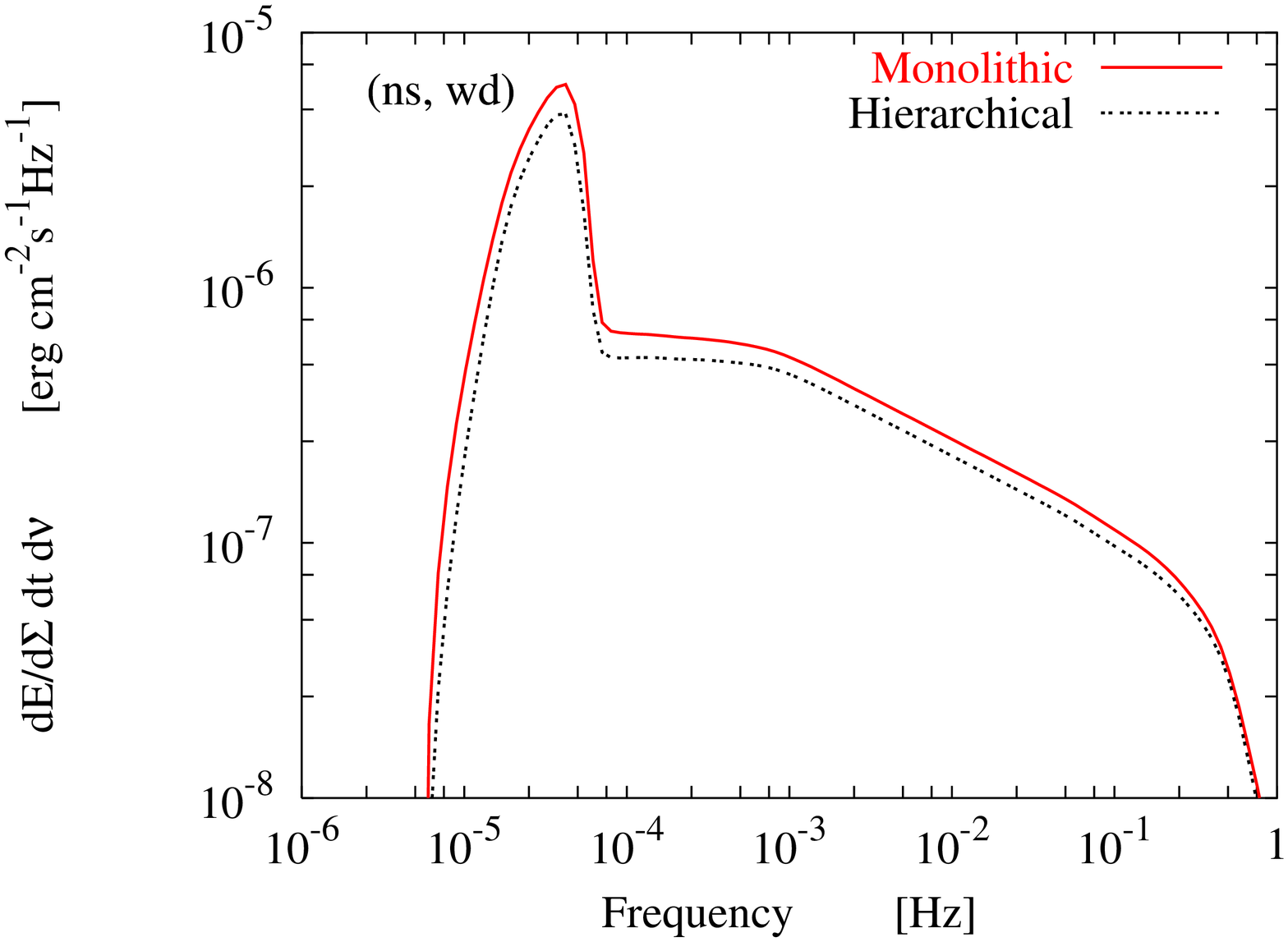,angle=270,width=7.cm}
\caption{The spectral energy density of the gravitational background
produced by various extragalactic populations of degenerate binaries in 
monolithic and hierarchical scenarios assuming a flat cosmological
background with zero cosmological constant.}  
\end{center}
\label{fig:bin_spectra}
\end{figure*}
%%%%%%%%%%%%%%%%%%%%%%%%%%%%%%%%%%%%%%%%%%%%%%%%%%%%%%%%%%%%%%%%%%%%%%%%%%%%%

The spectral energy densities are characterized by the presence of a sharp
maximum which, depending on the
binary population, has an amplitude spanning about two orders of magnitudes, 
in the frequency range $[10^{-5}-10^{-4}]$~Hz. In the following, we refer to
this part of the signal as 'primary' component. At higher frequencies, a
'secondary' component appears for all but (bh, bh) systems. The frequency
which marks the transition between primary and secondary components as
well as their relative amplitudes depend sensitively on the population.

The reason why (bh, bh) systems do not show a secondary component is that
this is entirely contributed by sources which merge before $z=0$. Conversely,
the low-frequency part of the spectrum is dominated by systems with 
merger-times larger than a Hubble time. 
These sources are observed at very low
frequencies because the value of the minimum frequency 
(which is emitted at formation, $z_f$) is set by the amplitude of the
merger-time [see eq.~(16)]. The larger is the merger-time, the smaller the
minimum frequency at which the in-spiral waves are emitted. Moreover, 
eq.~(6) shows that the flux emitted by each source decreases with frequency.
This explains the larger amplitude of primary components with respect to
secondary ones. 
For systems with merger-times larger than a Hubble time, the largest 
frequency is emitted at $z=0$ by binaries which form at $z_f \sim z_F$.
No contribution from such objects can be observed above this critical 
frequency and the primary component falls rapidly to zero. 

The amplitude of secondary components reflects the number of systems
with moderate merger-times. The maximum frequency which might be observed is
emitted by systems which are very close to their coalescence at $z=0$.
Since $\nu_{max}$ is larger for (ns, ns) than for (wd, wd), the secondary
component produced by double neutron stars extends up to $\sim10^2$~Hz.

It is interesting to note is that monolithic scenarios predict a 
maximum amplitude which is a factor $\sim$ 20-25\% larger than the
hierarchical case. This difference is much larger than what has been 
previously obtained for other extragalactic backgrounds (see \eg FMSI), 
indicating that
the energy density produced by extragalactic compact binaries is substantially
contributed by sources which form at redshifts $\gsim 1-2$. It is quite 
difficult to
unveil the origin of this effect because of the large number of parameters which
determine the appearance of the final energy density. However, a plausible explanation
might be that, depending on its specific time-delays $\tau_s$ and $\tau_m$, each
system emits the signal at redshifts which can be substantially smaller than the
formation redshift of the corresponding progenitor system. Thus, although the 
background signal is mostly emitted at low-to-intermediate redshifts, the sources
which produce these signals might have been formed at higher redshifts and 
reflect the state of the Universe at earlier times, when the differences among
hierarchical and monolithic scenarios are more significant.  
Comparing the different panels of Fig.~9, we conclude
that the background produced by (bh, bh) binaries has the largest amplitude
but it is concentrated at frequencies below $\sim 2\times 10^{-5}$~Hz. At
higher frequencies, which are more interesting from the point of view of 
detectability, the dominant contribution comes from (wd, wd) systems.
This is consistent with what has already been found for the galactic 
populations (Hils, Bender \& Webbink 1990).

>From the background spectrum it is possible to compute the closure density
$\Omega_{gw} \, h^2$ and the spectral strain amplitude of the signal $S_h$,
\beq
S_h(\nu)& =& \frac{2 G}{\pi c^3} \,\,\frac{1}{\nu^2} \,\,\frac{dE}{dSdtd{\nu}}({\nu}), \\
\Omega_{gw}(\nu)&=& \frac{\nu}{c^3 \rho_{cr}}\, \frac{dE}{dt dS
d\nu}(\nu).
\eeq 
The results are shown in Figs.~10 and~11
for all binary types within monolithic and hierarchical scenarios.
%%%%%%%%%%%%%%%%%%%%%%%%%%%%%%%%%%%%%%%%%%%%%%%%%%%%%%%%%%%%%%%%%%%%%%%%%%%%
\begin{figure*}
\begin{center}
\leavevmode
\psfig{figure=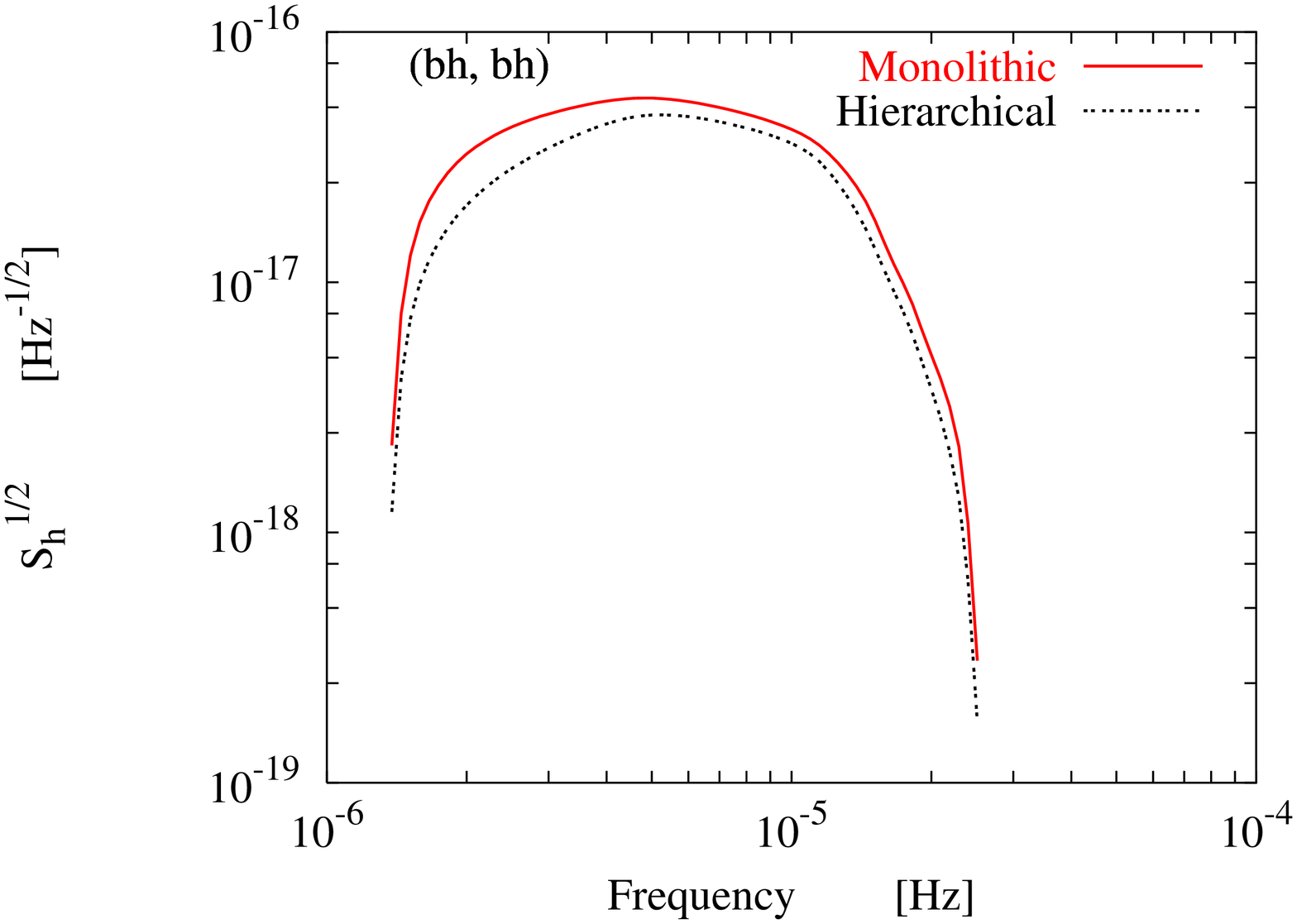,angle=270,width=7.cm}
\psfig{figure=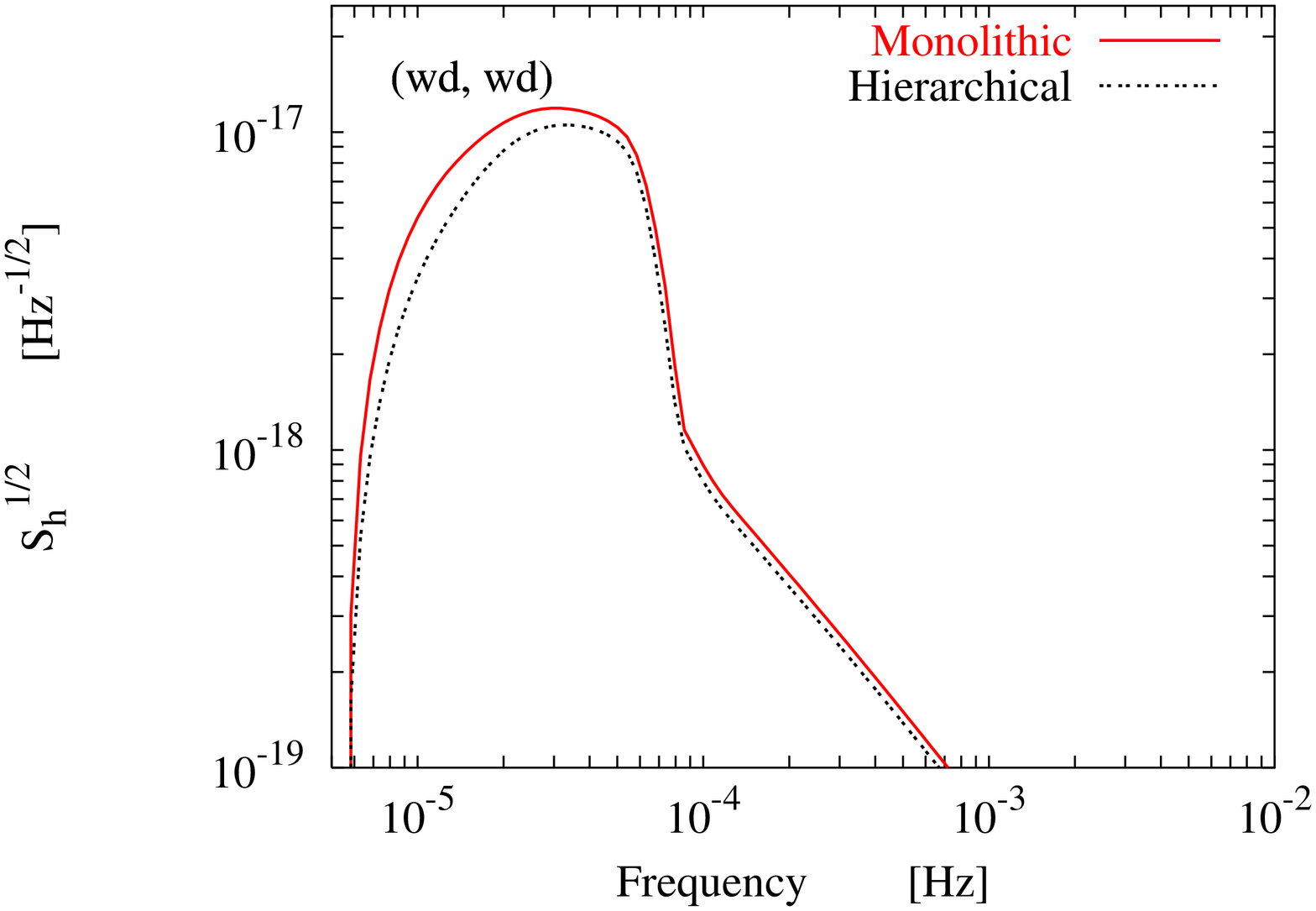,angle=270,width=7.cm}
\psfig{figure=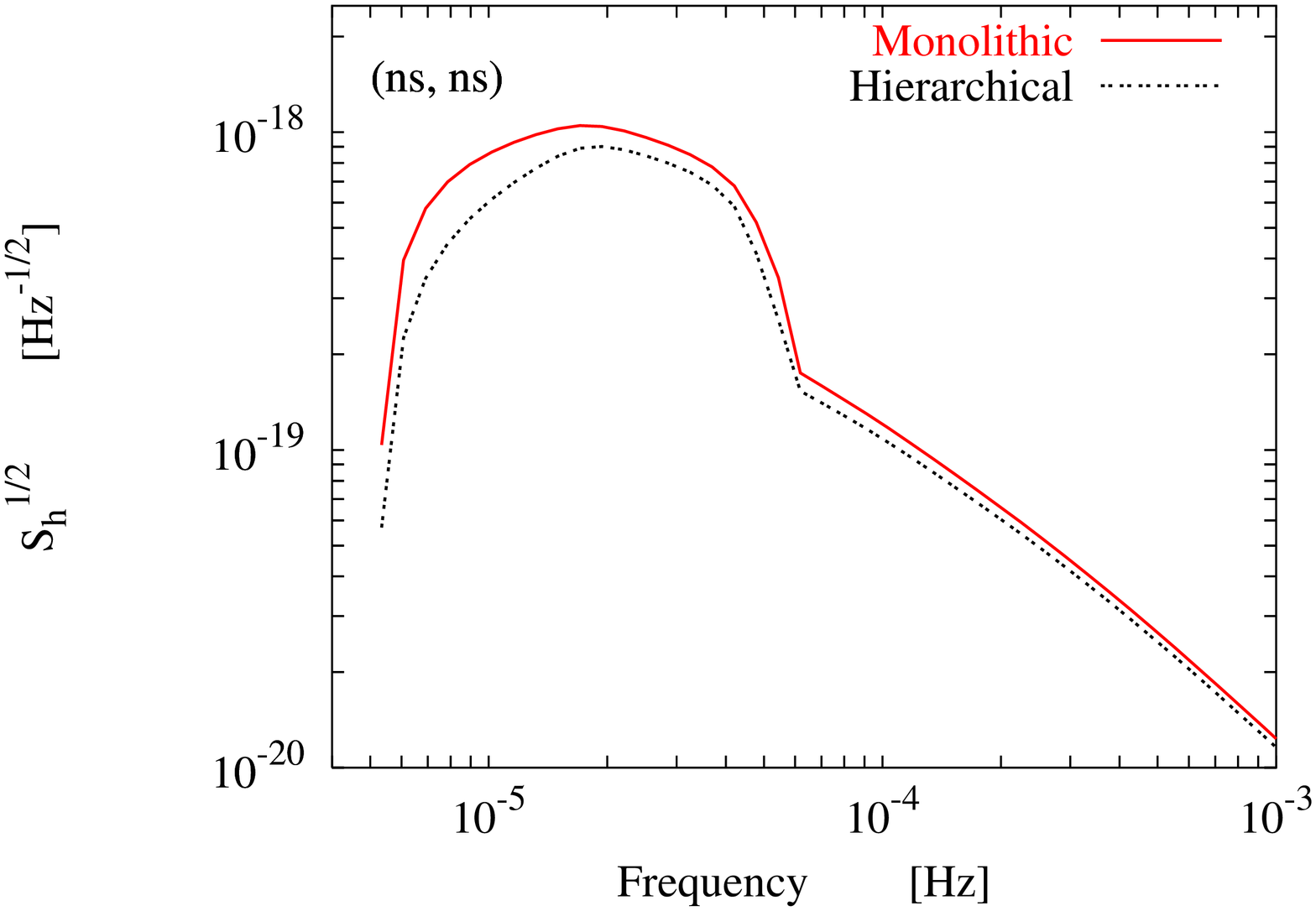,angle=270,width=7.cm}
\psfig{figure=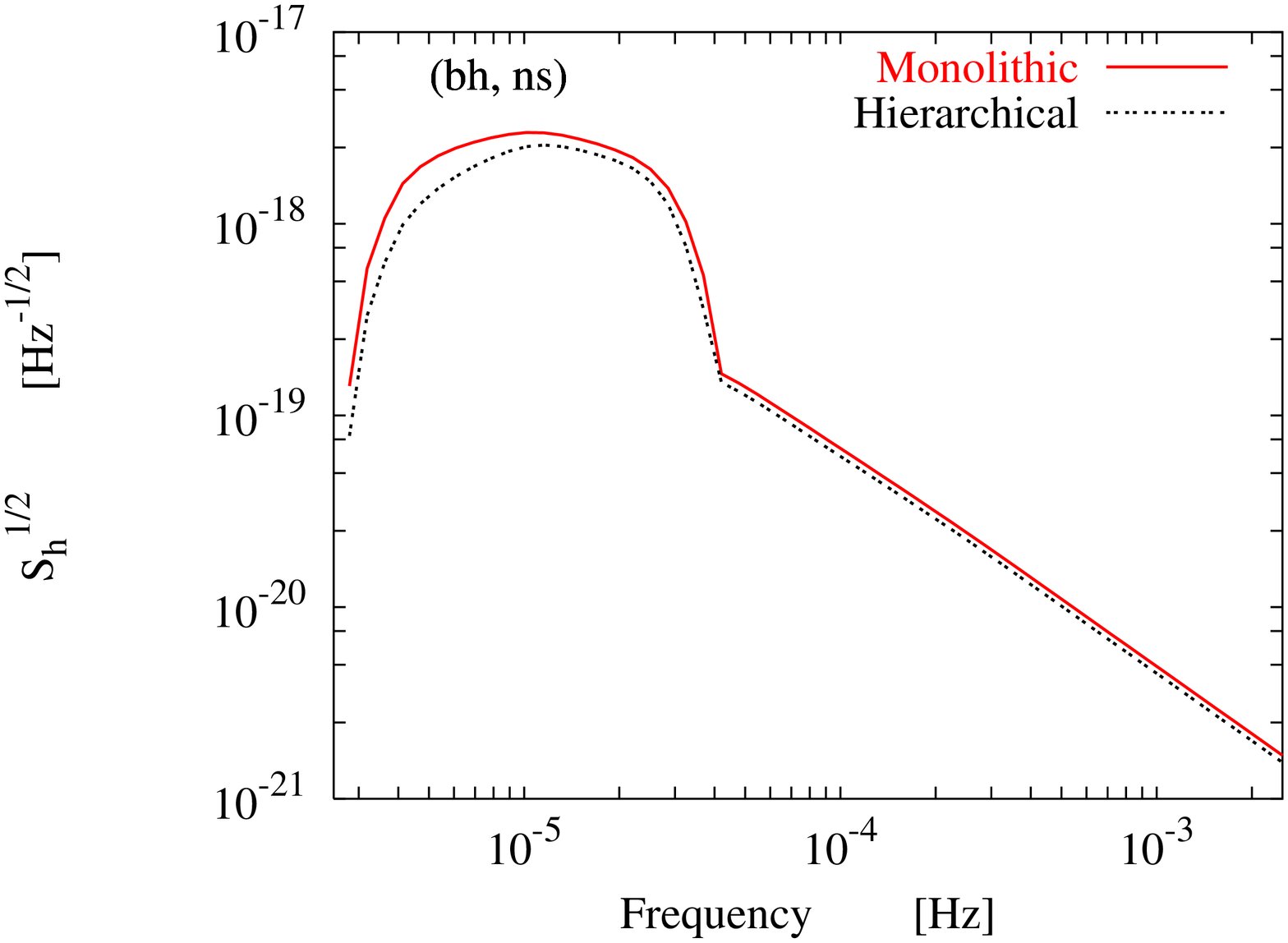,angle=270,width=7.cm}
\psfig{figure=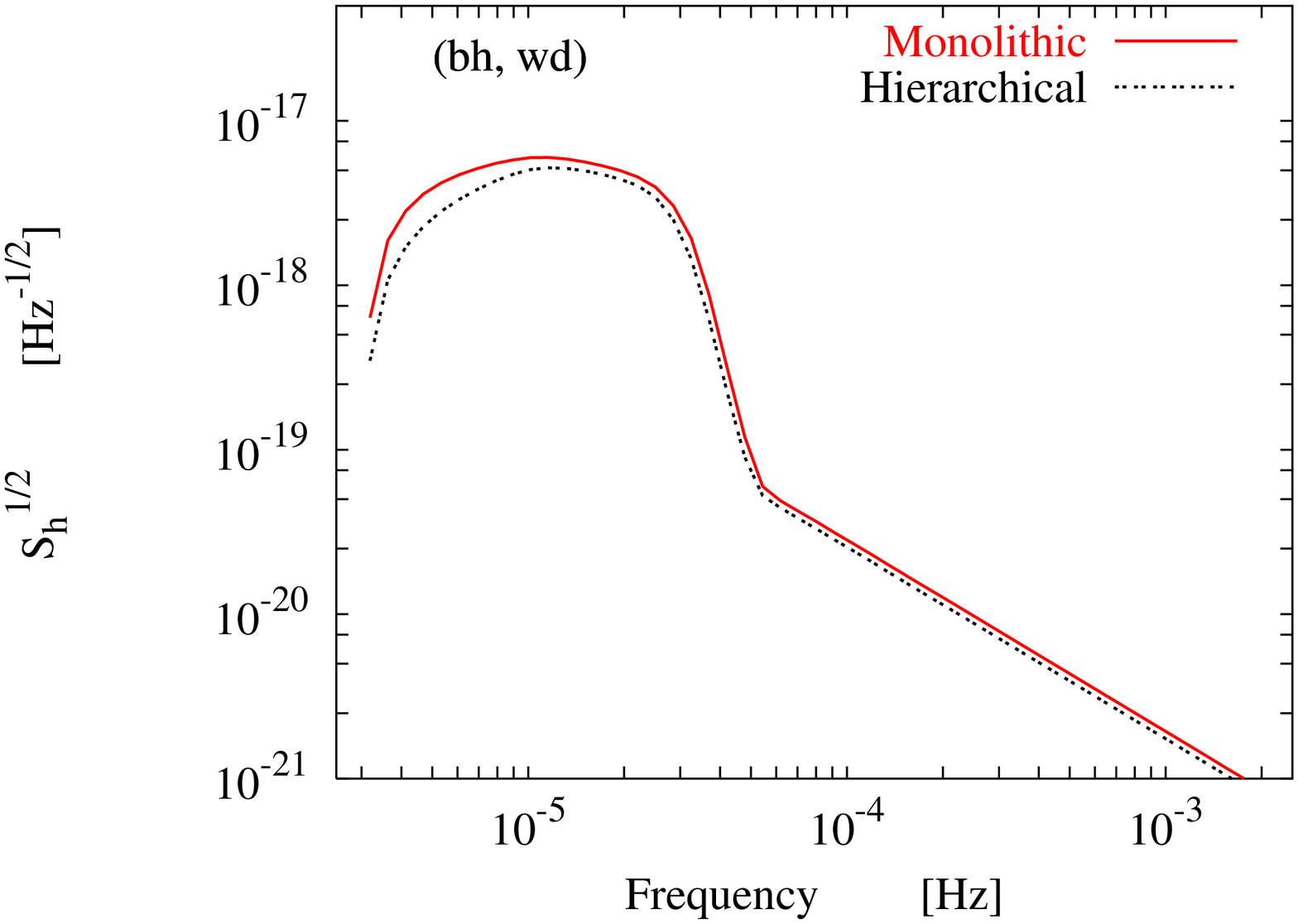,angle=270,width=7.cm}
\psfig{figure=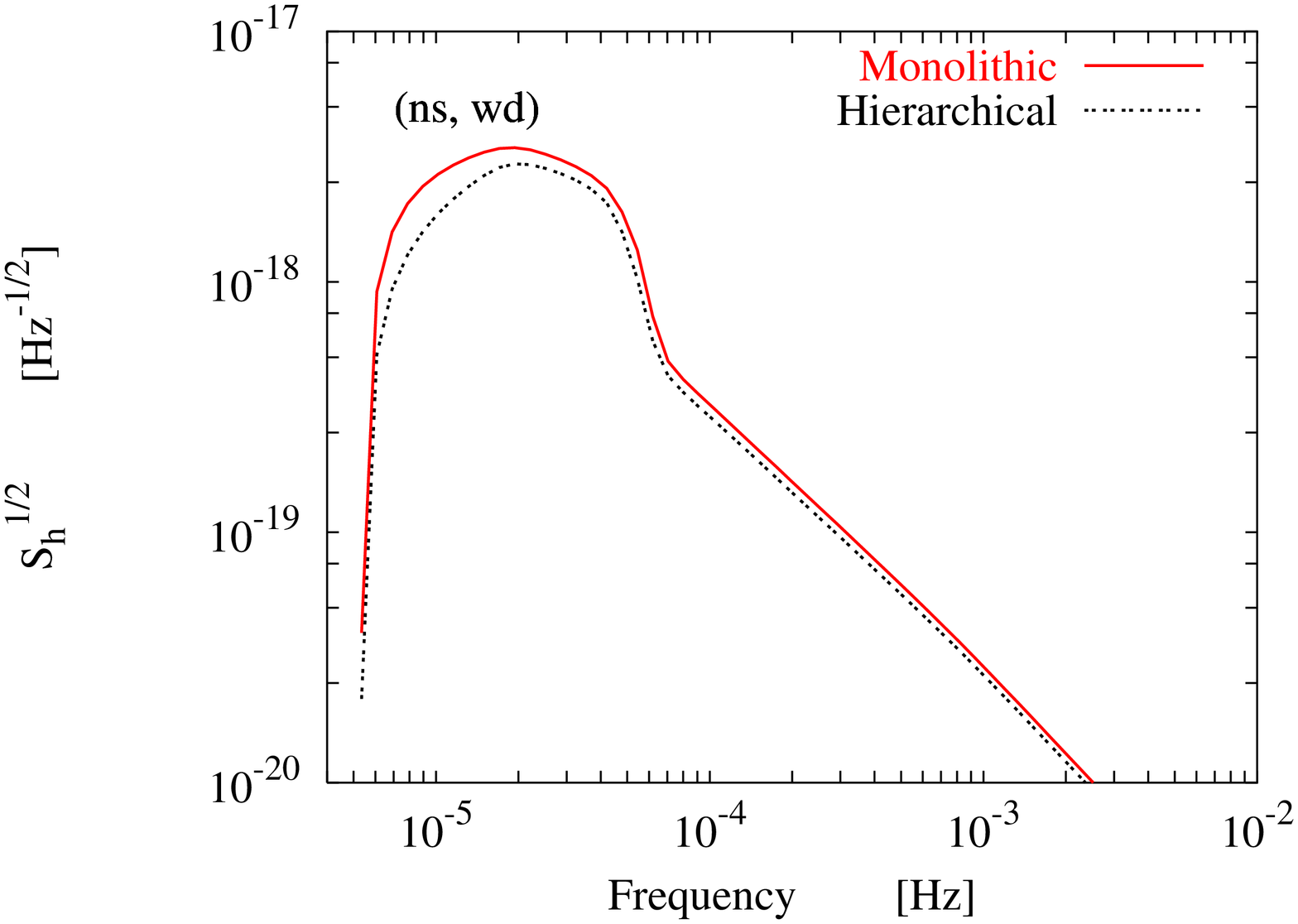,angle=270,width=7.cm}
\caption{The strain amplitude of the gravitational background
produced by various extragalactic populations of degenerate binaries in
monolithic and hierarchical scenarios assuming a flat cosmological background
model with zero cosmological constant.}  
\end{center}
\label{fig:bin_sh}
\end{figure*}
%%%%%%%%%%%%%%%%%%%%%%%%%%%%%%%%%%%%%%%%%%%%%%%%%%%%%%%%%%%%%%%%%%%%%%%%%%%%%

The strain amplitude of the backgrounds has a maximum amplitude
between $\sim 10^{-18} \, \mbox{Hz}^{-1/2}$ and
$\sim 5 \times 10^{-17}\,\mbox{Hz}^{-1/2}$ at frequencies in the interval
$[\sim 5 \times 10^{-6}-5 \times 10^{-5}]$~Hz.
The function $S_h$ is more sensitive to the low frequency part of the
energy density. 
Therefore, its shape reflects 
mainly the primary components of the corresponding 
energy density. In all but the (bh, bh) population,
it is evident the presence of a tail at frequencies above
the maximum which is the secondary component of the energy density: 
in the next section we compare this part of the
background signal with the LISA sensitivity to assess the possibility of
a detection. Still, it is clear that the prominent part of the background
signals produced by extragalactic populations of degenerate binaries 
could be observed with a detector sensitive to smaller frequencies than LISA.

Conversely, $\Omega_{gw} \, h^2$ is mostly dominated by secondary components.
We can compare the predictions for (bh, bh), (wd, wd) and (ns, ns)
systems. Contrary to what has been found for the spectral energy density
or for the strain amplitude of the signal, the largest $\Omega_{gw} \, h^2$
is produced by (ns, ns), as a consequence of 
the high amplitude of the  secondary component. 
In particular, no significant contribution
from the primary component appears.
For (wd, wd), instead, the contribution of the primary component
is relevant, although its amplitude is roughly half that of the secondary
component.
Finally, for (bh, bh) no secondary component is produced and thus the 
amplitude of the closure density is very low and at very low frequencies.
Mixed binary types have different properties, depending on the
relative importance of the above effect. 
For instance, (bh, wd) produce a secondary component but the amplitude is
so small to be comparable with that of the primary.

We stress that the value of $\nu_{max}$ is quite uncertain as it
defines the boundary between the early inspiral phase and the highly 
non-linear merger. Clearly, the more we get closer to this boundary, the
less accurate is the Newtonian description of the orbit as post-Newtonian
terms start to be relevant. Therefore, we believe that the most reliable 
part of the binary background signal is the low frequency part, \ie the
part which mostly contributes to the strain amplitude $S_h$.

%%%%%%%%%%%%%%%%%%%%%%%%%%%%%%%%%%%%%%%%%%%%%%%%%%%%%%%%%%%%%%%%%%%%%%%%%%%%
\begin{figure*}
\begin{center}
\leavevmode
\psfig{figure=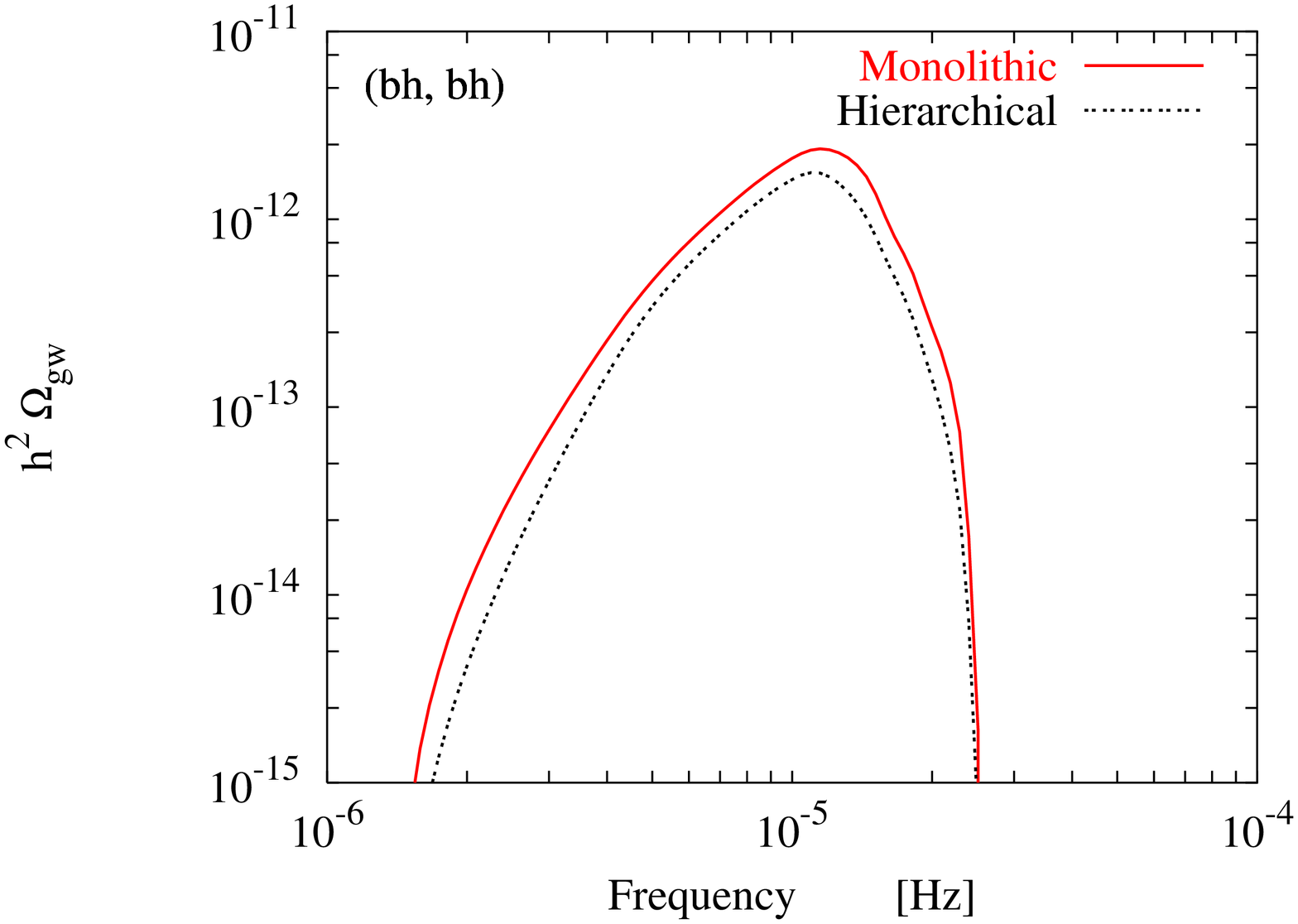,angle=270,width=7.cm}
\psfig{figure=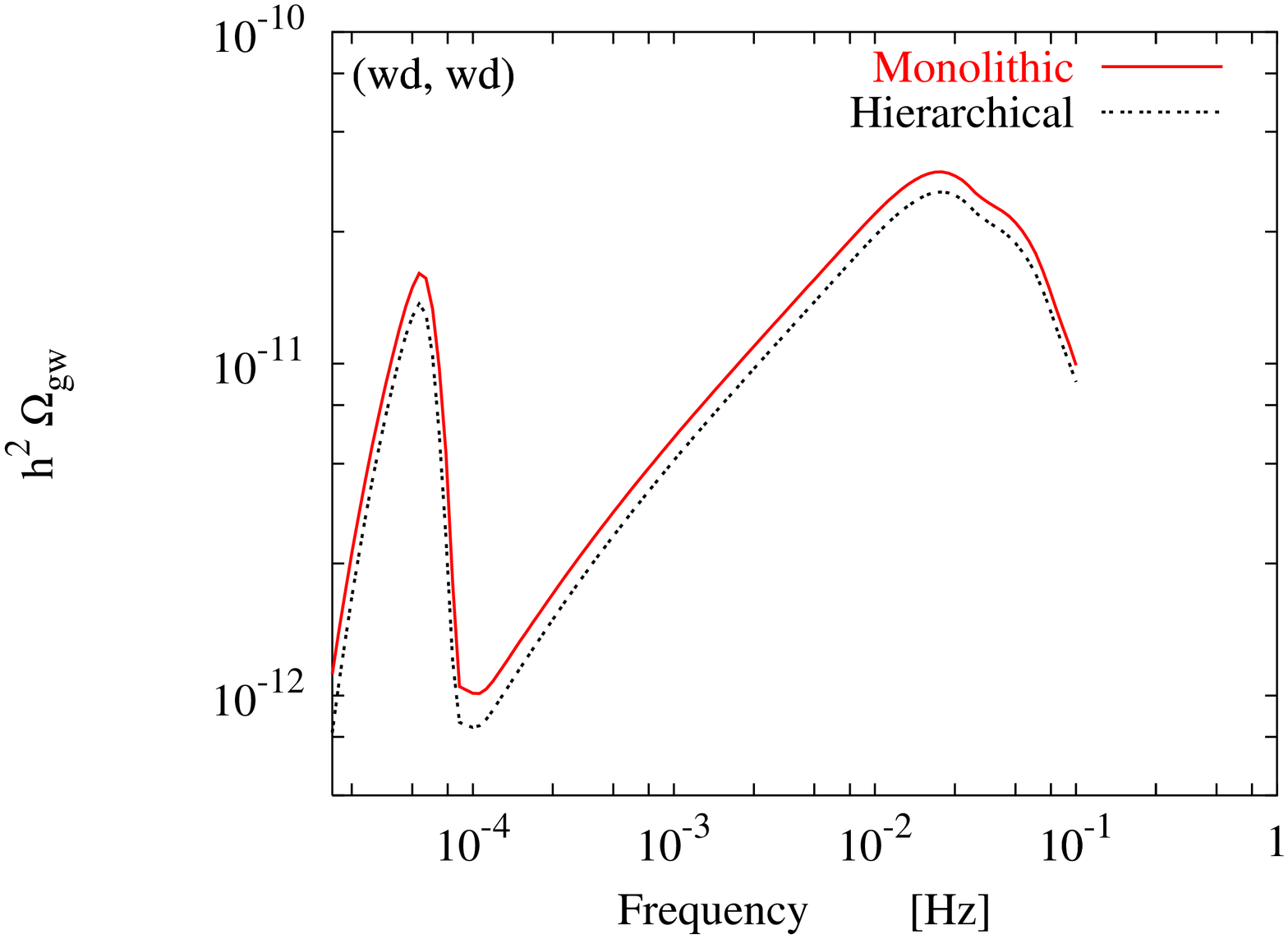,angle=270,width=7.cm}
\psfig{figure=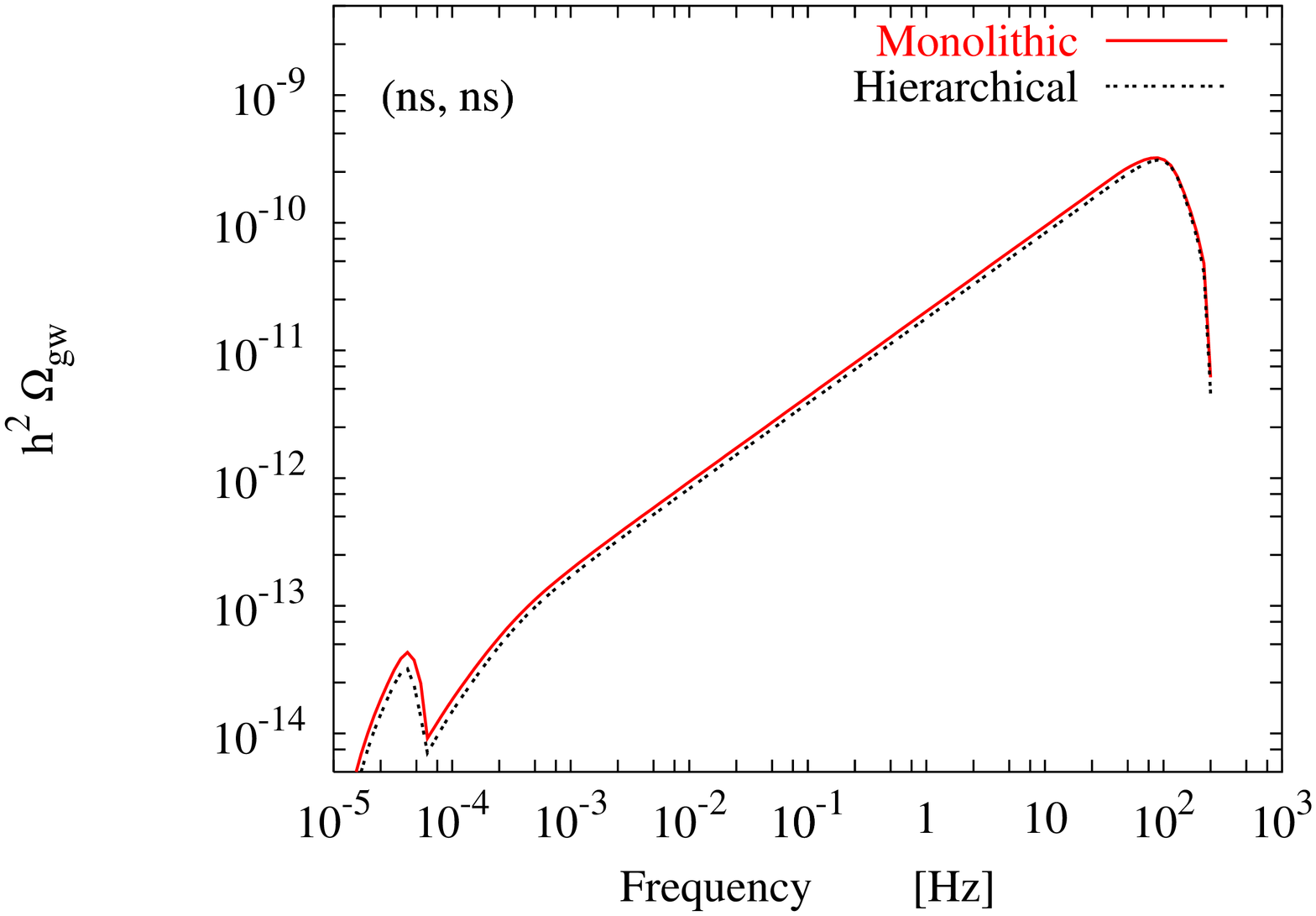,angle=270,width=7.cm}
\psfig{figure=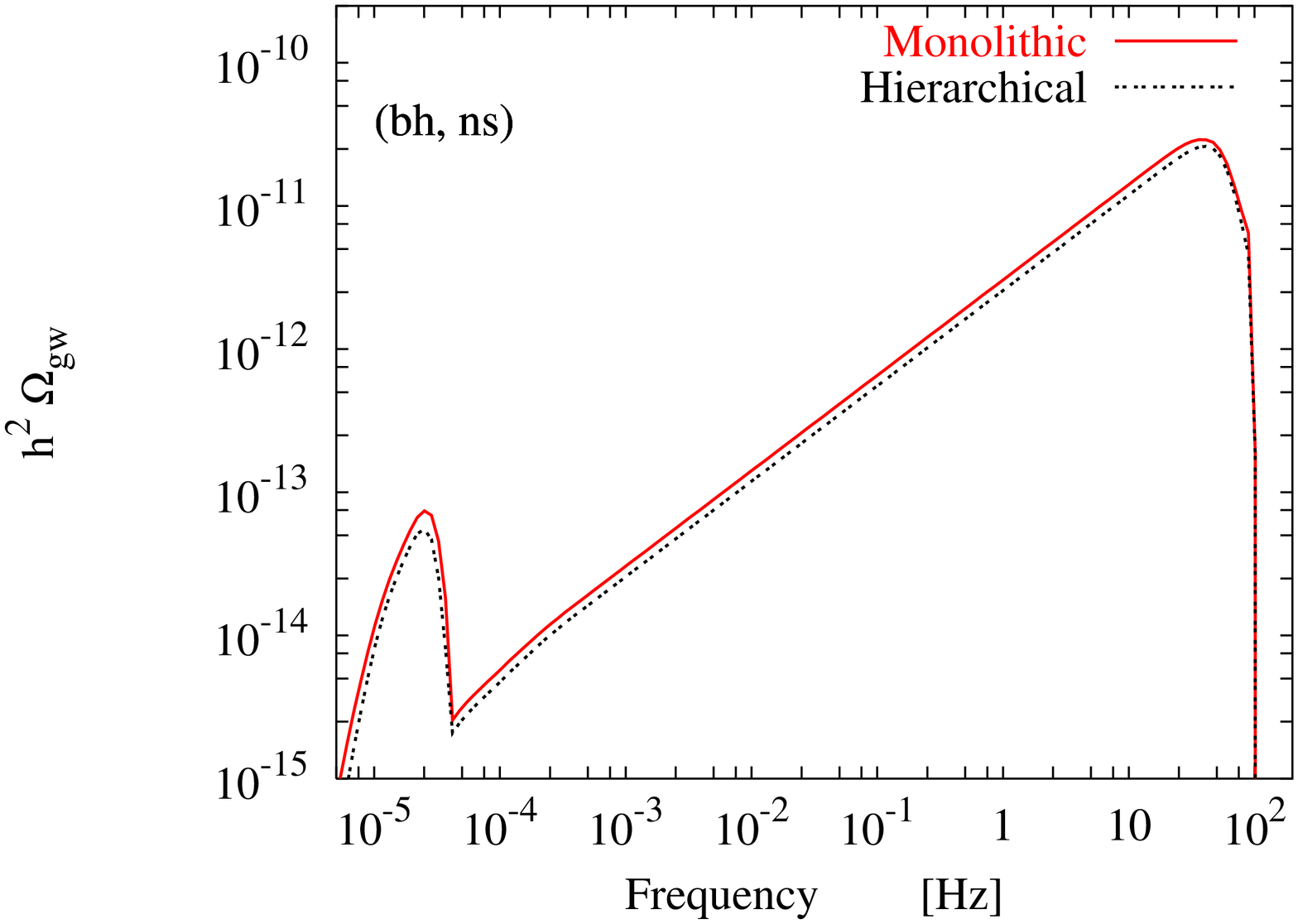,angle=270,width=7.cm}
\psfig{figure=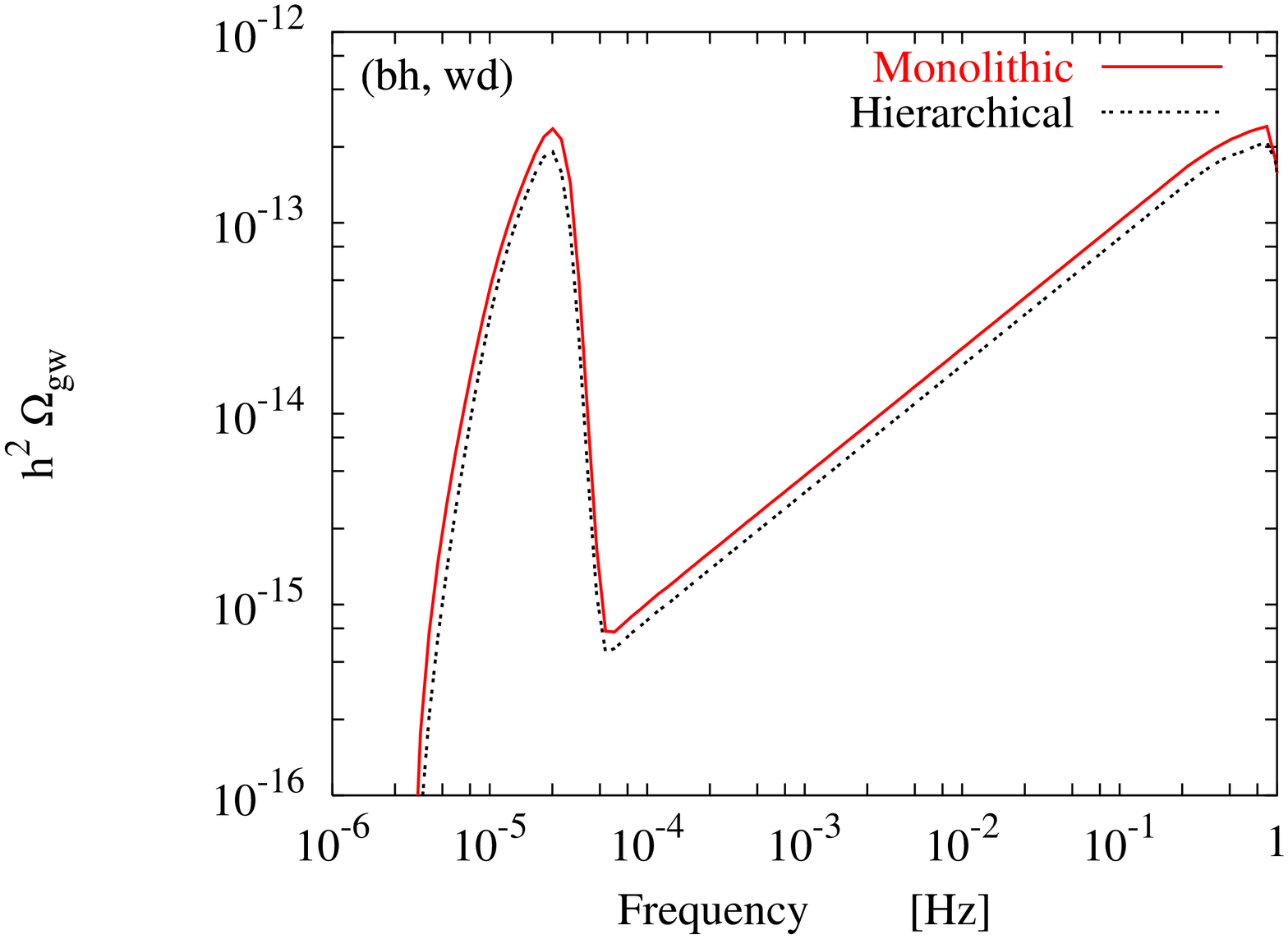,angle=270,width=7.cm}
\psfig{figure=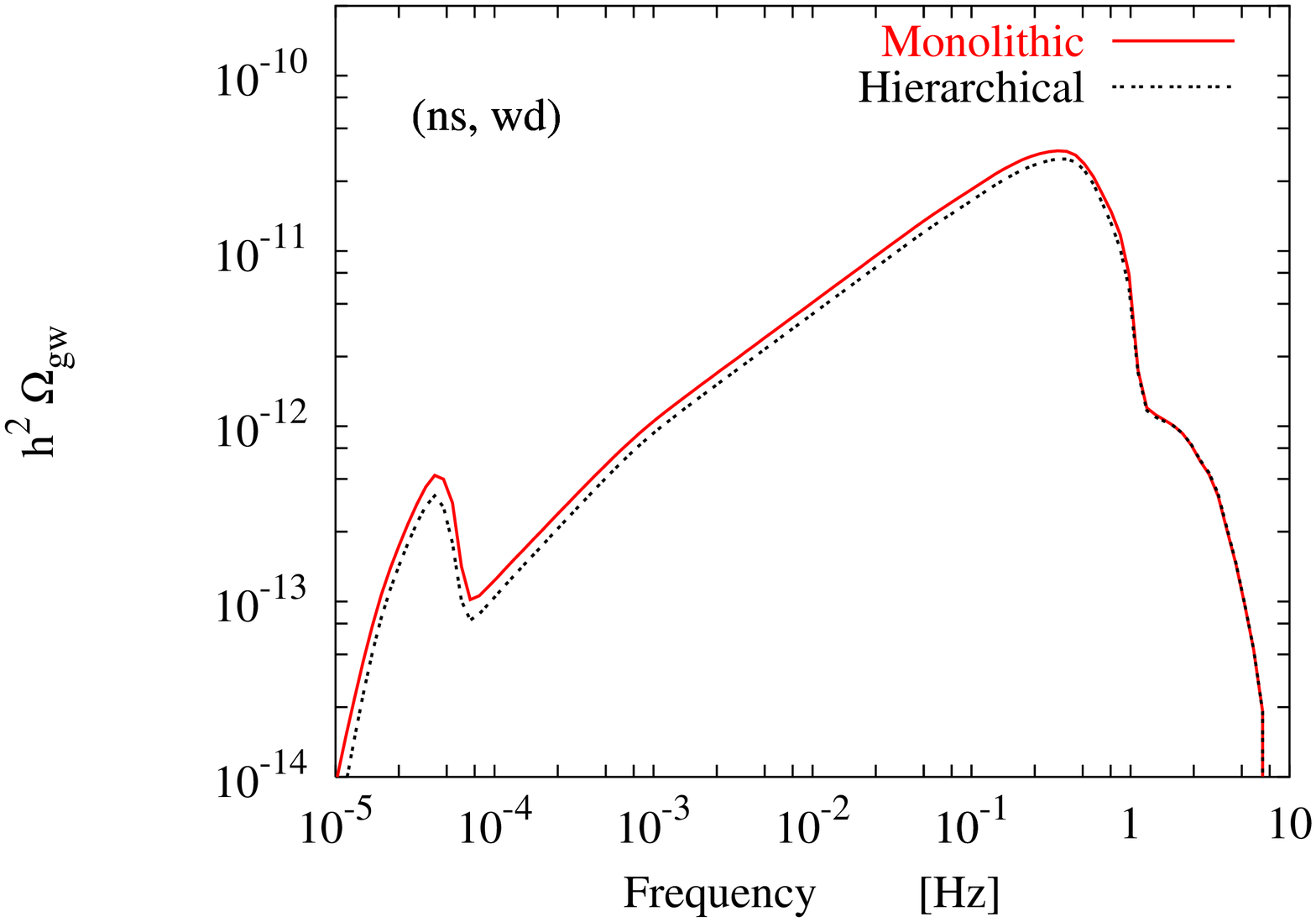,angle=270,width=7.cm}
\caption{The function $\Omega_{gw}\, h^2$ of the gravitational background
produced by various extragalactic populations of degenerate binaries in 
monolithic and hierarchical scenarios assuming a flat cosmological
background model with zero cosmological constant.}  
\end{center}
\label{fig:bin_omega}
\end{figure*}
%%%%%%%%%%%%%%%%%%%%%%%%%%%%%%%%%%%%%%%%%%%%%%%%%%%%%%%%%%%%%%%%%%%%%%%%%%%%%

\section{Confusion noise level and detectability by LISA}

To have some confidence in the detection of 
a stochastic gravitational background with LISA it is necessary 
to have a sufficiently large $\SNR$. The standard choice made by
the LISA collaboration is $\SNR=5$ which, in turn, yields a 
minimum detectable amplitude of a stochastic signal of 
(see Bender 1998 and references therein),
\be
h^2 \, \Omega_{gw}[\nu = 1\,\mbox{mHz}] \simeq 10^{-12}.
\ee
  
This value already accounts for the angle between the arms ($60^\circ$)
and the effect of LISA motion. It shows the remarkable sensitivity
that would be reached in the search for stochastic signals at low 
frequencies. Table~\ref{tbl:minima} shows that the backgrounds
generated by (wd, wd) and (ns, ns) 
extragalactic binary populations exceed this minimum
value and LISA might be able to detect these signals.

%%%%%%%%%%%%%%%%%%%%%%%%%%%%%%%%%%%%%%%%%%%%%%%%%%%%%%%%%%%%%%%%%%%%%
\begin{table}
\centering
\begin{tabular}{|c|cc|} \hline \hline 
$\nu=$1~mHz & (wd, wd) & (ns, ns) \\
\hline 
$\Omega_{gw}\,h^2$& $6 \times 10^{-12}$ & $1.1 \times 10^{-12}$ \\
\hline \hline
\end{tabular}
\caption{The values of the closure density at 1~mHz obtained for 
(wd, wd) and (ns, ns) extragalactic binary populations investigated. 
All these values
are larger than the minimum detectable value of $\Omega_{gw}\,h^2(1\mbox{mHz})$ predicted by the LISA team for a $\SNR=5$ and after 1 year of observation.}
\label{tbl:minima}
\end{table}
%%%%%%%%%%%%%%%%%%%%%%%%%%%%%%%%%%%%%%%%%%%%%%%%%%%%%%%%%%%%%%%%%%%%%%%%

\begin{figure}
\centerline{
\psfig{figure=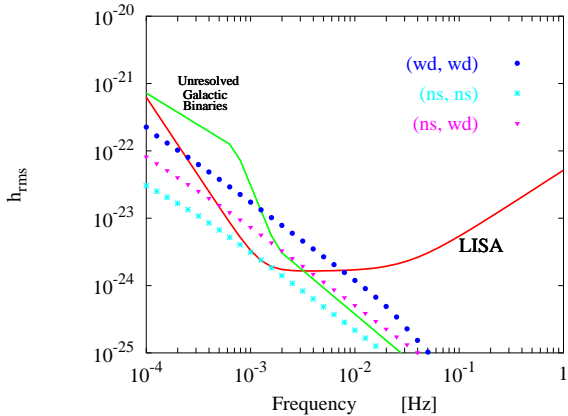,angle=270,width=8cm}}
\caption{The sensitivity of LISA to a stochastic background of gravitational
waves after one year of observation. The extragalactic backgrounds from
(wd, wd), (ns, ns), (ns, wd) and (bh, ns) systems 
might be observable at frequencies between $\sim 1$ and $\sim10$~mHz.}   
\label{fig:hrms}
\end{figure}

We plot in Fig.~\ref{fig:hrms} the predicted sensitivity of LISA to a stochastic
background after 1 year of observation (Bender 1998).
On the vertical axis it is shown $h_{\rms}$, defined as,
\be
h_{\rms} = [2\,\nu\, S_n(\nu)]^{1/2} \, \left(\frac{\Delta \nu}{\nu}\right)^{1/2}
\ee
where $S_n(\nu)$ is the predicted spectral noise density and the factor 
$(\Delta \nu/\nu)^{1/2}$ is introduced to account for the frequency
resolution $\Delta \nu = 1/T$ attained after a total observation time 
$T$.\\
On the same plot we show the equivalent $h_{\rms}$ levels predicted 
for different extragalactic binary populations and for the galactic 
population of close white dwarfs binary considered by Bender \& Hils
(1997).
%%%%%%%%%%%%%%%%%%%%%%%%%%%%%%%%%%%%%%%%%%%%%%%%%%%%%%%%%%%%%%%%%%%%%%%%%%%%
%%%%%%%%%%%%%%%%%%%%%%%%%%%%%%%%%%%%%%%%%%%%%%%%%%%%%%%%%%%%%%%%%%%%%%%%%%%%%

We see that the extragalactic backgrounds might be observable at frequencies
between $\sim 1$ and $\sim 10$~mHz. 

These background signals represent
additional noise components to the LISA sensitivity curve when
searching for signals from individual sources. 

\begin{figure}
\centerline{
\psfig{figure=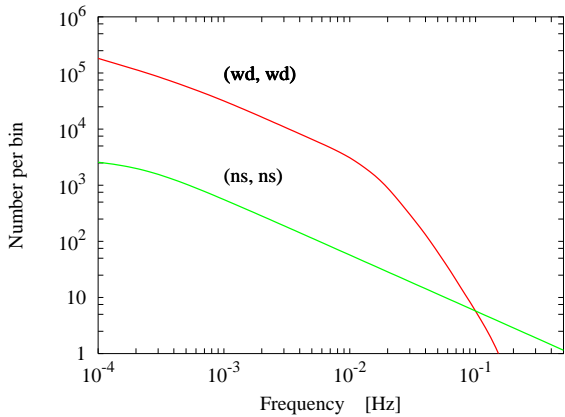,angle=270,width=8cm}}
\caption{The number of extragalactic (wd, wd) and (ns, ns) binaries per
resolution bin after a total observation time of 1yr.}   
\label{fig:numbin}
\end{figure}

In particular, backgrounds from unresolved astrophysical sources represent
a confusion limited noise. In fact, unless the signal emitted
by an individual source has a much higher amplitude, the background signal
prevents the individual source to be resolved. Clearly, the magnitude of this
effect depends on the frequency resolution of the instrument, \ie on the
observation time. The $h_{\rms}$ noise levels produced by extragalactic 
compact binaries shown in Fig.~\ref{fig:hrms} have been computed assuming
$T=1$~yr. For the same total observation time we show, 
in Fig.~\ref{fig:numbin}, the number of extragalactic (wd, wd) and (ns, ns) 
observed in each frequency resolution bin. At frequencies were these 
backgrounds might be relevant (between 1 and 10~mHz), the number of sources
per bin is $\gg 1$, representing a relevant confusion limited noise component.
The critical frequency above which the number of sources per bin is lower
than 1 occurs at $\sim 0.1$~Hz for (wd, wd) and outside LISA sensitivity 
window for (ns, ns). However, at these frequencies 
the dominant noise component is the instrumental noise.

%%%%%%%%%%%%%%%%%%%%%%%%%%%%%%%%%%%%%%%%%%%%%%%%%%%%%%%%%%%%%%%%%%%%%%%%%%%%
%%%%%%%%%%%%%%%%%%%%%%%%%%%%%%%%%%%%%%%%%%%%%%%%%%%%%%%%%%%%%%%%%%%%%%%%%%%%%

\section{Conclusions}

In this paper we have obtained estimates for the stochastic
background of gravitational waves emitted by cosmological populations of compact binary
systems during their early-inspiral phase.

Since we have restricted our investigation to frequencies well below the 
frequency emitted when each system approaches its last stable circular
orbit, we have characterized the single source emission 
using the quadrupole approximation. 

Our main motivation was to develop a simple method to estimate the 
gravitational signal produced by populations of binary systems at 
extragalactic distances. This method relies on three main pieces of 
information:
\begin{enumerate}
\item
the theoretical description of gravitational waveforms to 
characterize the single source contribution to the overall background    
\item
the predictions of binary population synthesis codes to characterize the 
distribution of astrophysical parameters (masses of the stellar components, 
orbital parameters, merger times etc.) among each ensemble of binary systems
\item 
a model for the evolution of the cosmic star formation history derived from
a collection of observations out to $z \sim 5$ to infer the evolution of the
birth and merger rates for each binary population throughout the Universe.
\end{enumerate}
As we have considered only the early-inspiral phase of the binary evolution,
our predictions for the resulting gravitational signals are restricted to the
low frequency band $10^{-5}-1$~Hz. The stochastic background signals produced
by (wd, wd) and (ns, ns) might be observable with LISA and 
add as confusion limited noise components to the LISA instrumental noise 
and to the signal produced by binaries within our own Galaxy. 
The extragalactic contributions are dominant at frequencies in the range 
$1-10$~mHz and limit the performances expected for LISA in the
same range, where the previously estimated sensitivity curve was attaining
its mimimum. 

We plan to extend further this preliminary study and to consider more
realistic waveforms so as to enter a frequency region interesting for
ground-based experiments.

\begin{figure}
\centerline{\psfig{figure=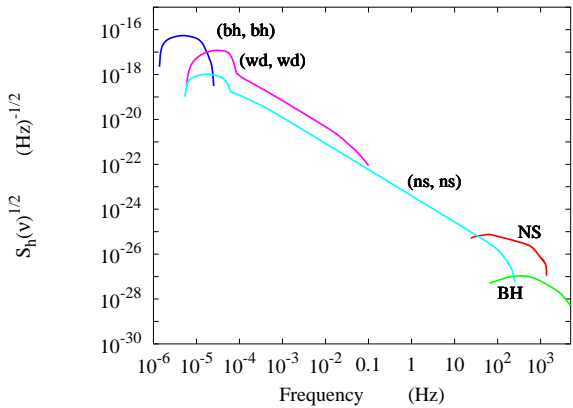,angle=270,width=8cm}}
\caption{The predicted strain amplitude of the stochastic backgrounds 
produced by extragalactic populations of gravitational sources. In the high
frequency band, we show the estimates for the background produced by 
rotating neutron 
stars via r-mode instability, and two possible signals emitted by populations
of massive stars collapsing to black holes (see text). In the 
low frequency band, we plot the background predicted  
for three different populations of binary systems.}
\label{fig:extragal}
\end{figure}

Finally, in Fig.~\ref{fig:extragal}
we show the spectral densities of the extragalactic backgrounds that have
been investigated so far. 
The high frequency band appears to be dominated by the stochastic
signal from a population of rapidly rotating neutron stars via the r-mode 
instability (see FMSII). For comparison, 
we have shown the overall signal emitted during
the core-collapse of massive stars to black holes (see FMSI). 
In this case, the amplitude
and frequency range depend sensitively on the fraction of progenitor 
star which participates to the collapse. 
The signal indicated with BH corresponds to the
conservative assumption that the core mass is $\sim 10 \%$ of the 
progenitor's (see FMSI).
Recent numerical simulations of core-collapse supernova
explosions (Fryer 1999) appear to indicate that for progenitor
masses $>40 \msun$ no supernova explosion occurs and the star directly 
collapses to form a black hole. The final mass of this core depends 
strongly on the relevance of mass loss caused by stellar winds 
(Fryer \& Kalogera 2000). 
If massive black holes are formed the resulting background would have
a larger amplitude and the relevant signal would be shifted towards 
lower frequencies, more interesting for ground-based interferometers
(Schneider, Ferrari \& Matarrese 1999).

In the low frequency band, we have plotted only the backgrounds produced by 
(bh, bh), (ns, ns) and (wd, wd) binaries because their signals largely 
overwhelm those from other degenerate binary types. 
%%%%%%%%%%%%%%%%%%%%%%%%%%%%%%%%%%%%%%%%%%%%%%%%%%%%%%%%%%%%%%%
%%%%%%%%%%%%%%%%%%%%%%%%%%%%%%%%%%%%%%%%%%%%%%%%%%%%%%%%%%%%%%%
%%%%%%%%%%%%%%%%%%%%%%%%%%%%%%%%%%%%%%%%%%%%%%%%%%%%%%%%%%%%%%%
%%%%%%%%%%%%%%%%%%%%%%%%%%%%%%%%%%%%%%%%%%%%%%%%%%%%%%%%%%%%%%%
 
We find that both in the low and in the high frequency band, 
extragalactic populations generate a signal which is comparable to and,
in some cases, larger than the
backgrounds produced by populations of sources within our Galaxy 
(Giazotto, Bonazzola \& Gourgoulhon 1997; Giampieri 1997; Postnov 1997;
Hils, Bender \& Webbink 1990; Bender \& Hils
1997; Postnov \& Prokhorov 1998; Nelemans, Portegies Zwart \& Verbunt 1999).
It is important to stress that even if future
investigations reveal that the amplitude of galactic 
backgrounds might be higher than presently conceived,
their signal could still be discriminated from that generated by
sources at extragalactic distance. In fact, the signal produced within the
Galaxy shows a characteristic amplitude modulation
when the antenna changes its orientation with respect to fixed stars
 (Giazotto, Bonazzola \& Gourgoulhon 1997; Giampieri 1997).

The same conclusions can be drawn when the extragalactic backgrounds are 
compared to the stochastic relic gravitational signals predicted by
some classical early Universe scenarios. 
The relic gravitational
backgrounds suffer of the many uncertainties which characterize our
present knowledge of the early Universe. According to
the presently conceived typical spectra, we find that their detectability
might be severely limited by the amplitude of the more recent astrophysical
backgrounds, especially in the high frequency band.

\section*{Acknowledgments}
We acknowledge Bruce Allen, Pia Astone, Andrea Ferrara, Sergio Frasca, Piero Madau and Lucia 
Pozzetti for useful conversations and fruitful insights in various aspects
of the work.

SPZ thank Gijs Nelemans and Lev Yungelson for discussions and
code developement. This work was supported by NASA through Hubble
Fellowship grant HF-01112.01-98A awarded (to SPZ) by the Space
Telescope Science Institute, which is operated by the Association of
Universities for Research in Astronomy, Inc., for NASA under contract
NAS\, 5-26555.  Part of the calculations are performed on the
Origin2000 SGI supercomputer at Boston University.  SPZ is grateful to
the University of Amsterdam (under Spinoza grant 0-08 to Edward
P.J. van den Heuvel) for their hospitality.

\label{lastpage}

\end{document}